\documentclass[preprint,12pt,authoryear]{elsarticle}

\usepackage{amssymb}

\usepackage{lineno}

\usepackage{lineno,hyperref}
\usepackage{placeins} 
\usepackage{scalefnt}
\usepackage{graphicx}

\usepackage{lscape}
\usepackage{ragged2e}
\modulolinenumbers[5]
\usepackage[%
    dvipsnames,
    svgnames,
    x11names,
    fixpdftex,
    table,
    xcdraw
]{xcolor}
\usepackage{rotating}
\usepackage{caption}
\usepackage{float}
\usepackage{rotating}

\journal{Journal of Systems and Software (JSS)}

\begin{document}

\begin{frontmatter}



\title{Psychometric Instruments in Software Engineering Research on Personality: Status Quo After Fifty Years}


\author[inst1]{Danilo Almeida Felipe}
\ead{dfelipe@inf.puc-rio.br}
\author[inst1]{Marcos Kalinowski}
\ead{kalinowski@inf.puc-rio.br}

\address[inst1]{Department of Informatics\\Pontifical Catholic University of Rio de Janeiro (PUC-Rio)\\
Rio de Janeiro - RJ, Brazil}

\author[inst2]{Daniel Graziotin}
\ead{daniel.graziotin@iste.uni-stuttgart.de}
\address[inst2]{University of Stuttgart, Institute of Software Engineering, Universitätsstraße 38 Stuttgart, Germany}
            
\author[inst3]{Jean Carlos Natividade}
\ead{jeannatividade@puc-rio.br}
\address[inst3]{Department of Psychology\\Pontifical Catholic University of Rio de Janeiro (PUC-Rio)\\
Rio de Janeiro - RJ, Brazil}

\begin{abstract}
\noindent
\textbf{Context:} Although software development is a human activity, Software Engineering (SE) research has focused mostly on processes and tools, making human factors underrepresented. This kind of research may be improved using knowledge from human-focused disciplines. An example of missed opportunities is how SE employs psychometric instruments.
\\
\textbf{Objective:} Provide an overview of psychometric instruments in SE research regarding personality and provide recommendations for adopting them.
\\
\textbf{Method:} We conducted a systematic mapping to build an overview of instruments used within SE for assessing personality and reviewed their use from a multidisciplinary perspective of SE and social science.
\\
\textbf{Results:} We contribute with a secondary study covering fifty years of research (1970 to 2020). One of the most adopted instruments (MBTI) faces criticism within social sciences, and we identified discrepancies between its application and existing recommendations. We emphasize that several instruments refer to the Five-Factor Model, which despite its relevance in social sciences, has no specific advice for its application within SE. We discuss general advice for its proper application.
\\
\textbf{Conclusion:} The findings show that the adoption of psychometric instruments regarding personality in SE needs to be improved, ideally with the support of social science researchers. We believe that the review presented in this study can help to understand limitations and to evolve in this direction.
\\
\end{abstract}




\end{frontmatter}


\newcommand{\eg}{\textit{e.g.}, }
\newcommand{\ie}{\textit{i.e.}, }
\newcommand{\Ie}{\textit{I.e.}, }
\newcommand{\cf}{\textit{cf.}, }
\newcommand{\Cf}{\textit{Cf.}, }

\newcommand{\rqmap}{Which psychometric instruments have been applied in SE research regarding personality?}

\newcommand{\rqsurvey}{How do social sciences researchers perceive the adoption of psychometric instruments regarding personality in SE research?}

\newcommand{\rqmapA}{What are the objectives of the studies?}

\newcommand{\rqmapB}{What are the limitations faced by the use of psychometric instruments reported in the studies by the authors?}

\newcommand{\rqmapC}{To which SE constructs are those psychometric instruments related?}

\newcommand{\rqmapD}{Which types of research do the studies refer to?}

\newcommand{\rqmapE}{Which types of empirical studies have been conducted?}


\section{Introduction}
\label{cha:Introduction}

Software Engineering (SE) activities are primarily performed by humans. However, many empirical studies have only focused on proposing new methods and technologies to support SE activities leaving human and social factors behind them underexplored \citep{feldt_towards_2008}, impeding a more holistic view of the area.

Behavioral Software Engineering (BSE), proposed by \cite{lenberg_behavioral_2015}, is the body of knowledge of SE research that attempts to understand human aspects related to the activities of software engineers, software developers, and other stakeholders. The area has been the subject of recent research in the SE domain. Nevertheless, because BSE is relatively immature, some approaches adopted misled researchers mainly by not properly combining SE research with social sciences backgrounds, such as psychology, to address human factors \citep{graziotin_affect_2015,graziotin2022psychometrics}.

Furthermore, measurement activities are an essential part of empirical software engineering research and most quantitative studies. In empirical studies, researchers and readers must possess a high degree of confidence regarding how the results of measuring resources can be interpreted in valid and reliable ways. These resources can be personnel, hardware, or software for an activity or process \citep{wohlin_experimentation_2012}

BSE research has encouraged the use of psychometric instruments as support to the understanding of human factors in a more systematic way \citep{feldt_towards_2008, lenberg_behavioral_2015}. In its turn, the Psychoempirical Software Engineering proposed by \cite{graziotin_affect_2015} deals with ``denoting research in Software Engineering with established theory and measurements from psychology''. A problem addressed by the authors is the misuse of theoretical backgrounds of psychology, such as assuming a certain theory as the only truth in the research foundation; and also the improper use of psychometric instruments, some not validated within the psychology area or used to evaluate wrong human factors.

\cite{graziotin2022psychometrics} has further highlighted the importance of properly adopting psychometric instruments in SE. They introduced  introductory guidelines for SE researchers, including an example of psychometric validation with the R language. They covered topics such as operationalizing psychological constructs, item pooling, item review, pilot testing, item analysis, factor analysis, statistical property of items, reliability, validity, and fairness in testing and test bias. They report that ``to improve the quality of behavioral research in SE, studies focusing on introducing, validating, and then using psychometric instruments need to be more common''. Our study echoes this request.

We notice that information on \textit{whether and how psychometric instruments are adopted in SE research} remains vague and dispersed in many studies. As far as we know, one study has partially synthesized this knowledge concerning a specific construct, personality\footnote{Personality is one of the most studied concepts in BSE research as pointed by \cite{lenberg_behavioral_2015}}: a period of forty years (1970 to 2010) about personality in SE research is mapped by \cite{cruz_forty_2015}. Still, there is only a brief discussion and characterization of the instruments and their use in the software engineering context (\eg education and pair programming), missing a critical assessment. This status quo remains unchanged more than ten years later and deserves to be challenged.

There is also a need to find out whether SE research over the years has been adopting these instruments coherently with well-known recommendations. Although some authors have outlined this on a smaller scale \citep{mcdonald_who_2007, usman_use_2019}, a large-scale study has not been done to get a big picture.

In order to synthesize this knowledge in a structured manner, \textbf{our objective is to present an overview and reflections on the use of psychometric instruments in SE research on personality}\footnote{Hereafter we refer to psychometric instruments related to personality simply as ``psychometric instrument(s)''.}. 

We intend to consolidate findings on the use of psychometric instruments in SE research. Therefore, we searched for new studies using psychometric instruments by applying an update strategy to a systematic mapping on personality-related SE research by \cite{cruz_forty_2015}. It is noteworthy that our study does not strictly correspond to an update of the previous study, as we focus specifically on the use of psychometric instruments. Given that the population of studies to analyze the instruments is the same, \ie primary studies reporting SE research on personality, new studies within this same population can be identified by following the guidelines for the search strategy to update secondary studies by \cite{wohlin_guidelines_2020}. 

More specifically, we classify and discuss the objective of the studies, reported limitations on their use, SE constructs related to the psychometric instruments, the type of research, and empirical evaluations. Additionally, we discuss aspects of the most used instruments under the lens of literature guidelines and involved a social science researcher active with psychometric and personality-related research (the last author) to provide critical feedback to the SE community.

Our main contribution concerns updating evidence on personality-related SE research in order to cover fifty years of research, with a particular focus on psychometric instruments. We carefully assessed the need for an update of the evidence \citep{mendes_when_2020}, and we followed guidelines on the search strategy to update systematic literature studies \citep{wohlin_guidelines_2020}. The previous mapping study covered forty years of research and identified 90 research papers \citep{cruz_forty_2015}. Applying the search strategy led us to identify 106 additional papers published within the ten subsequent years (2011 to 2020). More specific contributions related to the psychometric instruments include:

\begin{itemize}

    \item Outlining the use of psychometric instruments in SE research on personality over fifty years.

    \item Observing remaining discrepancies between one of the most adopted instruments (MBTI) within recent SE research and existing recommendations in the literature. We also observed several studies using the Five-Factor Model, while specific advice on how to apply this model within the SE domain is missing. We discuss general advice from social science research.
    
    \item Identifying the most common objectives of studies employing psychometric instruments and the limitations of their application as reported by the authors.
   
    \item Relating the use of psychometric instruments within recent SE research to theoretical SE constructs, aiming at providing a better understanding on how such instruments are used within the context of \textit{actors} applying \textit{technologies/interventions} to perform \textit{activities} on \textit{software systems}.
    
    \item Summarizing the type of research and the empirical evaluations in SE research employing psychometric instruments.

\end{itemize}

The remainder of this paper is organized as follows. Section \ref{sec:Previous Work} provides background on Behavioral Software Engineering, common personality models, and secondary studies that address personality in SE, focusing on psychometric instruments. Section \ref{sec:SMP} presents the systematic mapping protocol, derived research questions, update strategy for getting new evidence on psychometric instruments, and documented execution of the protocol. Section \ref{sec:Results} presents the results organized by research questions. Section \ref{sec:results-discussion} provides a discussion on the evolution of psychometric instruments over the whole fifty-year period and reviews their use based on literature guidelines. Section \ref{sec:results-threats} discusses threats to validity. Finally, Section \ref{sec:Conclusion} presents the concluding remarks, limitations, and future work.


\section{Background and Related Work}
\label{sec:Previous Work}

This section introduces the theoretical foundation and related work to this study. To approach the theoretical foundation, we describe the broader context of this study (Behavioral Software Engineering), provide a definition of personality, and describe frequently used personality models in SE. Regarding related work, we focus on personality studies in SE, emphasizing existing secondary studies.

\subsection{Behavioral Software Engineering (BSE)}
BSE is defined as ``the study of cognitive, behavioral and social aspects of software engineering performed by individuals, groups or organizations'' \citep{lenberg_behavioral_2015}. It involves dealing with existing relationships between SE and disciplines from social sciences, such as \textit{work and organizational psychology}, the \textit{psychology of programming}, and \textit{behavioral economics} to get a broader understanding of SE practices.

Although software is developed by humans, for a long-time, SE research has focused intensively on the technical aspects (such as processes and tools) and less on human and social aspects \citep{feldt_towards_2008}.

BSE introduces several constructs called BSE concepts. When operationalized in empirical research, they could provide insights to researchers and practitioners. Also, with adequate knowledge adopted from other disciplines, such as social sciences, it is possible to better understand the software engineer's practices, as a human, in the execution of their activities. It is worth mentioning that BSE is restricted to software engineers and stakeholders, and not human aspects related to the use of the software \citep{lenberg_behavioral_2015}.

Still in the context of BSE, the authors present a definition for the body of knowledge research described earlier and also conducted a systematic literature review based on the definition. The findings report lack of research in some SE knowledge areas (\eg requirements, design, and maintenance) and rare collaboration between SE and social science researchers. A list of 55 BSE concepts and respective units of analysis is raised and detailed. Concepts such as \textit{cognitive style}, \textit{job satisfaction}, \textit{communication}, and \textbf{\textit{personality}} are seen as more frequently studied, while concepts such as \textit{intentions to leave} are underexplored.

\subsection{Personality: Adopted Definition and Common Models}

In the present study, we focused on mapping the literature on BSE with a restricted scope in personality, observed as one of the most studied concepts and the one with the most significant relationship with other concepts \citep{lenberg_behavioral_2015}. Despite being a human factor with different definitions, we need to adopt a consistent one with the consolidated literature in SE. A deeper review and discussion are not our scope and goal. However, we need to adopt definitions to support our decisions. For personality, we rely on the following definition used in \cite{cruz_forty_2015}:

\begin{quotation}
Personality is generally viewed as a dynamic organization, inside the person, of psychophysical systems that create the person’s characteristic patterns of behavior, thoughts, and feelings. \cite{ryckman_theories_2012} defined personality as ``the dynamic and organized set of characteristics possessed by a person that uniquely influences his or her cognitions, motivations, and behaviors in various situations''. We use these definitions because they are general enough to allow the inclusion of studies covering a wide range of personality theories and research methods. [...] The dispositional perspective encompasses the traits and types theory, which is one of the most used theories in organizational psychology 
\citep{anderson_handbook_2001} and in studies on personality in software engineering.
\end{quotation}

Complementarily, \cite{barroso_influence_2017} summarize personality models commonly used to identify personality traits in SE in dispositional perspective. Those models are the \textbf{Myers-Briggs Type Indicator} and the \textbf{Five-Factor Model}.

The Myers-Briggs Type Indicator (MBTI) is a model based on Jung's theory of personality types adapted by Isabel Myers and Katharine Briggs in a personality inventory, with the purpose of identifying dominant individual preferences over four dichotomous dimensions \citep{myers_mbti_1998}:
    
    \begin{itemize}
    
    \item \textit{I-E dimension}: refers to the way that an individual directs their energy towards the world. Introversion (I) directs to the inner world of experiences, ideas, and internal experiences (imaginative world). Extraversion (E) to the outer world of people and objects (the real world).
    
    \item \textit{S-N dimension}: refers to functions or the way of an individual's perception work. Sensing (S) people tends to rely on what can be perceived by the five senses; on the other side, iNtuitive (N) people rely on patterns and relationships.
    
    \item \textit{T-F dimension}: refers to the processes of judging and make conclusions. Thinking (T) people tend to perform logical and objective impartial analysis. Feeling (F) people highlight personal or social values, in a harmonic way.
    
    \item \textit{J-P dimension}: is an extended dimension based on Jung's theory, this refers to the way that an individual prefers to deal with the outside world. Judging (J) people prefer to be decisive using the judging processes (T-F dimension). Perceiving (P) people prefer to be more spontaneous using perception processes (S-N dimension).
    
    \end{itemize}
    
Given the dimensions described earlier, an individual can be placed in one of 16 possible combinations of personality (INTP, ESFJ, ISFJ, and so on).

Unlike the MBTI, the Five-Factor Model (FFM) does not provide a classification of people into types. Instead, this model allows assessing levels of latent traits underlying behaviors, levels that vary from person to person on a continuum (at least, in theory). The FFM has proved to be one of the most consensual perspectives in describing personality structure  \citep{mccrae_five-factor_2008}. This model describes personality based on broad five factors, which are latent traits with a high explanatory scope on behaviors: \textit{Extroversion}, \textit{Agreeableness}, \textit{Conscientiousness}, \textit{Neuroticism}, and \textit{Openness to experience}. 

The FFM emerged from strictly empirical approach and gained prominence in the 1990s. Starting from the idea that language could provide data to characterize personality traits (see \cite{john_lexical_1988}), researchers, using factorial analysis techniques, found five-factor structures to explain the correlations between the items in their instruments \citep{john_history_2021}. The model proved even more consolidated when researchers, using personality instruments developed under other theories, also found a five-factor structure (\eg \cite{costa_jr_catalog_1988, lanning_dimensionality_1994}).

Table \ref{tab:background-rw} provides descriptions by \cite{feist_theories_2012} of the Five-Factor Model, in which a person tends to show higher or low scores in characteristics by each broad dimension.

\begin{table}[H]
\centering
\caption{Description of the Five-Factor Model broad dimensions.}
\label{tab:background-rw}
\resizebox{.65\textwidth}{!}{%
\begin{tabular}{cl|l}
\multicolumn{1}{l}{\textbf{}} &
  \multicolumn{1}{c|}{\textbf{Higher scores}} &
  \multicolumn{1}{c}{\textbf{Lower scores}} \\ \hline
\textbf{Extraversion} &
  \begin{tabular}[c]{@{}l@{}}affectionate\\ joiner\\ talkative\\ fun loving\\ active\\ passionate\end{tabular} &
  \begin{tabular}[c]{@{}l@{}}reserved\\ loner\\ quiet\\ sober\\ passive\\ unfeeling\end{tabular} \\ \hline
\textbf{Neuroticism} &
  \begin{tabular}[c]{@{}l@{}}anxious\\ temperamental\\ self-pitying\\ self-conscious\\ emotional\\ vulnerable\end{tabular} &
  \begin{tabular}[c]{@{}l@{}}calm\\ even-tempered\\ self-satisfied\\ comfortable\\ unemotional\\ hardy\end{tabular} \\ \hline
\textbf{Openness to experience} &
  \begin{tabular}[c]{@{}l@{}}imaginative\\ creative\\ original\\ prefers variety\\ curious\\ liberal\end{tabular} &
  \begin{tabular}[c]{@{}l@{}}down-to-earth\\ uncreative\\ conventional\\ prefers routine\\ uncurious\\ conservative\end{tabular} \\ \hline
\textbf{Agreeableness} &
  \begin{tabular}[c]{@{}l@{}}softhearted\\ trusting\\ generous\\ acquiescent\\ lenient\\ good-natured\end{tabular} &
  \begin{tabular}[c]{@{}l@{}}ruthless\\ suspicious\\ stingy\\ antagonistic\\ critical\\ irritable\end{tabular} \\ \hline
\textbf{Conscientiouness} &
  \begin{tabular}[c]{@{}l@{}}conscientious\\ hardworking\\ well-organized\\ punctual\\ ambitious\\ persevering\end{tabular} &
  \begin{tabular}[c]{@{}l@{}}negligent\\ lazy\\ disorganized\\ late\\ aimless\\ quitting\end{tabular}
\end{tabular}%
}
\end{table}

Often the Five-Factor Model is refereed in the literature as Big Five [Model]. \cite{barroso_influence_2017} points out an existing distinction between then in theoretical basis, causality, and measurement between the two models, distinguishing the Five-Factor as a derivation of the Big Five, in which the latter assumes that personality traits are important for social interaction. At the same time, the former is a model that provides causes and contexts. A deeper review of personality models is beyond our scope and we consider the Five-Factor Model as the main one.

Given the definition of personality and an overview of common personality models, in the next section we highlight secondary studies on the dispositional perspective of personality. Some of them were captured by our secondary study protocol described in Section \ref{sec:SMP}.

\subsection{Personality and Psychometrics in Software Engineering}
\label{sec:background-personalitySE}

According to \cite{michell_measurement_1999}, “psychometrics is concerned with theory and techniques for quantitative measurement in psychology and social sciences” (Michell, 1999 \textit{apud} Feldt et al., 2008). In addition, \cite{feldt_towards_2008} state that ``[...] in practice, this often means the measurement of knowledge, abilities, attitudes, emotions, \textbf{personality}, and motivation''. The use of psychometric instruments in SE is encouraged, especially in empirical research, as a way to emphasize hitherto unexplored human factors and to help understand how they affect the research landscape \citep{feldt_towards_2008}. This is our focus in this study, given the importance of measurement activities in empirical research.

In \cite{barroso_influence_2017}, personality models utilized in 21 papers are mapped onto three main ones: Myers-Briggs Type Indicator (MBTI), Five-Factor Model, and Big Five. The study covers the period of 2003 to 2016 and includes peer-reviewed publications in the IEEE, ACM, and Elsevier digital libraries. In addition to the personality models used, inconclusive findings are also identified on the influence of software engineers' personalities on professional activities. However, there is no mapped information about the psychometric instruments that operationalize these models.

As far as we know, \cite{mcdonald_who_2007} is the first study that brings to attention on the use of psychometric instruments in SE research, beyond providing guidelines for the use of two of them (MBTI and 16PF). In addition, is highlighted that one of the authors is from social sciences and a certified professional regarding these instruments.

\cite{cruz_forty_2015} performed a systematic mapping on personality in SE research using the dispositional perspective. In addition to reporting on the most common SE topics addressed, such as \textit{education} and \textit{extreme programming}, they reported which personality tests (a.k.a psychometric instruments) were most commonly used, which resembles this study. However, the authors only reported brief information on personality-related psychometric instruments without deep discussion about them regarding their use in SE. They provide a valuable list of instruments and relate them to some SE topics, but their scope did not include analyzing how these instruments were applied and the limitations reported by the authors.

\cite{usman_use_2019} investigate ethical topics raised by \cite{mcdonald_who_2007} on the adoption of MBTI-based tests in a sample of 8 studies obtained in the final set compiled by \cite{cruz_forty_2015} published after 2007, and complemented with 7 studies returned in string-based search on Scopus\footnote{https://www.scopus.com/} in the years of 2016 and 2017, totaling a sample of 15 studies. Their results indicate that the use of psychometric instruments in SE is inadequate. The authors found problems in all of the analyzed studies, including the reliability and validity of MBTI (there are different versions of this instrument). The authors also highlight possible causes, such as not exploring literature guidelines and lack of collaboration with social science researchers. However, the study reported is initial and limited to analyzing only the use of MBTI in a small sample of studies.

Still, \cite{graziotin_understanding_2015} claims that the use of psychometric instruments should be cautious, in addition to the proper theoretical background used. The authors then propose the \textit{Psychoempirical Software Engineering} that aims ``to denote research in SE with proper theory and measurement from psychology''. In the same study, the authors provide broader steps when adopting psychometric instruments in SE research and exemplify scenarios using the \textit{affect} construct. Nonetheless, the steps initially do not cover personality.

Building on the previous study,~\cite{graziotin2022psychometrics} conducted a survey of the psychometric theory literature guided by \textit{The Standards for Educational and Psychological Testing}. They repackaged the synthesized knowledge as introductory guidelines for SE researchers, including an example of psychometric validation with the R language. The reviewed topics were operationalizing psychological constructs, item pooling, item review, pilot testing, item analysis, factor analysis, statistical property of items, reliability, validity, and fairness in testing and test bias. The paper provides guidelines that encourage a culture change in SE research toward the adoption of established methods from psychology.


\section{Systematic Mapping Protocol}
\label{sec:SMP}

Systematic mapping is a method to build a classification scheme of an area providing a visual summary of the state of research in a structured way \citep{petersen_systematic_2008}. It aims at providing an auditable and replicable process with minimal bias.

This section describes each step of our research method based on guidelines in the literature. Subsection \ref{sec:smp-rqs} introduces the mapping goal and research questions. Subsection \ref{sec:smp-search} describes the search strategy for collecting new evidence. Subsection \ref{sec:smp-studysel} presents the study selection criteria and discusses quality assessment, and Subsection \ref{sec:smp-def} presents the Data Extraction Form and our classification scheme. Concluding, Subsection \ref{sec:applying-smp} documents how the mapping protocol was started.

\subsection{Mapping Goal and Research Questions}
\label{sec:smp-rqs}

Our systematic mapping aims at providing an overview of the use of psychometric instruments in SE research. To guide our investigation, and to obtain an overview of the state-of-the-art, trends and gaps, we describe the following Research Questions (RQs) as follows:

\begin{itemize}
    \item \textbf{RQ1: }\rqmap    
    \item \textbf{RQ2: }\rqmapA
    \item \textbf{RQ3: }\rqmapB
    \item \textbf{RQ4: }\rqmapC
    \item \textbf{RQ5: }\rqmapD
    \item \textbf{RQ6: }\rqmapE
\end{itemize}

In the following section, the search strategy is presented. It was developed by the first author and reviewed by the second one.

\subsection{Search Strategy}
\label{sec:smp-search}

\subsubsection{The Need to Update Evidence on Psychometric Instruments}

This mapping started in a traditional way of conducting secondary studies (string-based search in digital libraries with snowballing steps), according to consolidated literature \citep{petersen_systematic_2008, petersen_guidelines_2015, kitchenham_guidelines_2007, mourao_investigating_2017}. Later, new guidelines emerged, and we noticed that they could help conduct our study \citep{mendes_when_2020, wohlin_guidelines_2020}, given the awareness we had about comprehensive secondary studies on human factors in SE \citep{cruz_forty_2015, lenberg_behavioral_2015}.

We defined \cite{cruz_forty_2015} as a candidate study to be considered as we wanted to start our immersion into BSE using a narrower scope, focused on personality, to allow a comprehensive overview and a focused critical assessment. \cite{cruz_forty_2015} identified 90 studies published within a time range of forty years, whereas \cite{lenberg_behavioral_2015} defined 55 BSE concepts (\textit{e.g.}: \textit{personality}, \textit{job satisfaction}, \textit{communication}, etc.) in a large scale study that considered 250 papers. The narrower focus by \cite{cruz_forty_2015} would also allow us to apply our update search strategy to get new evidence (discussed in Subsection \ref{sec:smp-update}) with reasonable efforts.

We also argue that personality is a BSE concept presented in \cite{lenberg_behavioral_2015} as one of the most studied together with others (such as \textit{group composition}, \textit{communication}, and \textit{organizational culture}). We believe that updating evidence regarding psychometric instruments of \cite{cruz_forty_2015} study yields significant results regarding our objective described in Section \ref{cha:Introduction}, which is also within BSE's scope. Thus, we decided to update the mapping study by \cite{cruz_forty_2015}.

\begin{figure}[h]
\centering
\includegraphics[width=11cm]{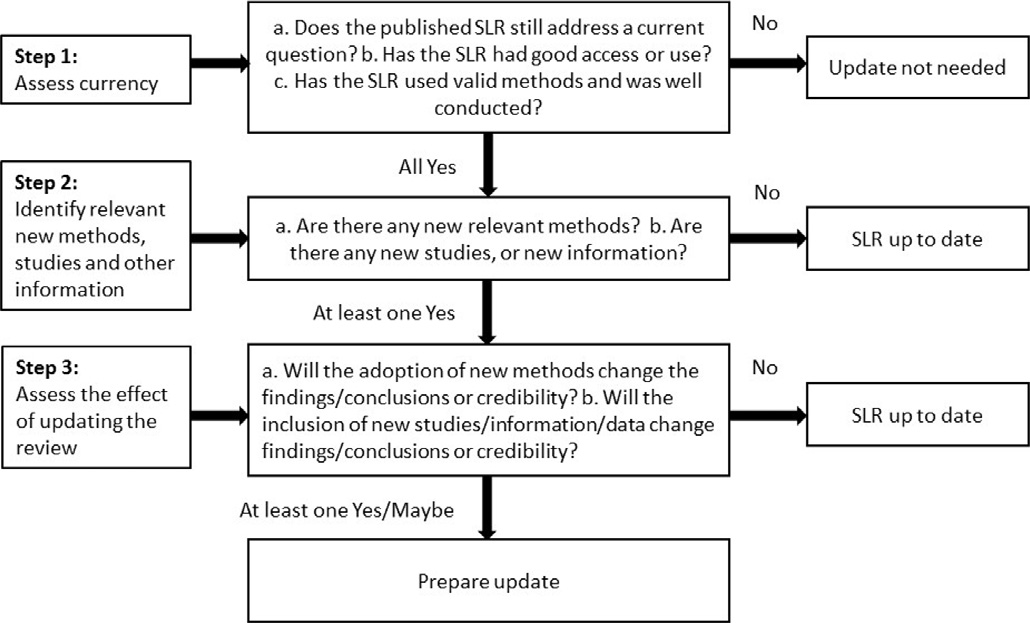}
\caption{3PDF framework recommended by \cite{mendes_when_2020}.}
\label{fig:3pdf-framework}
\end{figure}

We used the 3PDF decision framework recommended by \citep{mendes_when_2020} with a specific focus on identifying the need of updating evidence on psychometric instruments, using as a basis the evidence contained on this subject in the mapping study conducted by \cite{cruz_forty_2015}. We conducted the evaluation process by answering the seven 3PDF questions, which are listed and answered hereafter (in Steps 1.a. to 3.b.) and illustrated in Figure \ref{fig:3pdf-framework}.

\textbf{Step 1.a.: }\textit{Does the published SLR still address a current question?}    
Yes. Regarding the currency of the evidence on psychometric instruments contained in the specific literature mapping by \cite{cruz_forty_2015}, their study has been used to subsidize information on psychometric instruments in dozens of current studies on personality, including studies recently published in some of the main software engineering journals. For instance, a recent paper by \cite{russo_gender_2022}, addressing gender differences in personality traits of software engineers, refers to \cite{cruz_forty_2015} to emphasize that software engineering has adopted several instruments to measure personality. As additional examples, \cite{glasauer2023prevent} and \cite{amin_impact_2020} also cite the paper by \cite{cruz_forty_2015} to back information on the main psychometric instruments being used.

Considering that the study by \cite{cruz_forty_2015} only covers until 2010, arguments on the main instruments being used would ideally need to be backed by more recent evidence. Hence, there is a need to update the status quo on psychometric instruments to cover the entire half a century.

\textbf{Step 1.b.: }\textit{Has the SLR had good access or use?} Yes. Like \cite{mendes_when_2020}, we used in the cut-off point the same yearly average citation value of 6.82 documented by \cite{garousi_highly-cited_2016} to consider a paper for good access or use. In August of 2020, \cite{cruz_forty_2015} had a yearly average citation value of 30.8 in Google Scholar. Furthermore, we identified recent papers indicating that part of its access and use refers to backing up evidence on psychometric instruments.

\textbf{Step 1.c.: }\textit{Has the SLR used valid methods and was it well-conducted?} 
Yes. Regarding the methods, \cite{cruz_forty_2015} present an extension of preliminary results published previously in \cite{cruz_personality_2011} with improvements, such as a refined search string increasing the sensitivity and coverage; adding backward snowballing steps; review of research questions and extended presentation of results.
Finally, the authors present clear steps in their mapping protocol and are based on well-recognized guidelines for conducting secondary studies in software engineering \citep{kitchenham_guidelines_2007}.

\textbf{Step 2.a.: }\textit{Are there any new relevant methods?} Yes. We identified new guidelines to efficiently search for evidence to update secondary studies in software engineering \citep{wohlin_guidelines_2020}. As \cite{cruz_forty_2015} used valid methods (Step 1.c.) to identify evidence including psychometric instruments, we decided that we could confidently use their findings as a basis for updating such evidence using the identified guidelines. 

\textbf{Step 2.b.: }\textit{Are there any new studies or new information?} Yes. The papers included in the original mapping study had each a considerable number of citations in a preliminary verification in Google Scholar. There is more than a decade of evidence on psychometric instruments not incorporated by \cite{cruz_forty_2015}, which covered only until 2010.

\textbf{Step 3.a.: }\textit{Will the adoption of new methods change the findings, conclusions or credibility?} Yes, potentially. We adopted a new method concerning the mapping protocol and addressed different research questions to get a big picture of psychometric instruments in software engineering research, which we believe generates new and important findings. \cite{cruz_forty_2015} has a relevant research question on psychometric instruments, but there is little discussion of its results beyond listing and counting frequencies.

\textbf{Step 3.b.: }\textit{Will the inclusion of new studies/information/data change findings, conclusions or credibility?} Yes. Regarding new potential findings, we had prior knowledge of a series of studies \cite{graziotin_affect_2015, graziotin_consequences_2017, graziotin_what_2018} used as control papers as a strategy to ensure good literature coverage in our protocol. These studies are not covered by \cite{cruz_forty_2015} because they were published later. They discuss the use of psychometric instruments in SE on the theoretical basis of other areas (such as social sciences and psychology), which can support SE research in general.

It is noteworthy that, although we identified the need for an update, our goal and research questions differ, as we focused on psychometric instruments. The authors of the original study focused on characterizing the SE research on personality. Hence, we share the same population of studies to be analyzed. Although their study also mapped the instruments by answering the research question ``What personality tests are administered in the studies, and to what type of participants (professionals or students)?'', they did not openly provide a detailed spreadsheet of extracted data. This limited us to work only with the information reported in their paper.

\subsubsection{Strategy to Collect New Evidence}
\label{sec:smp-update}

We adopted the guidelines proposed by \cite{wohlin_guidelines_2020} as a strategy to search for new evidence based on an existing secondary study. They are the following:

\begin{itemize}
    \item \textbf{Use a seed set containing the original secondary study and its included primary studies:} \cite{cruz_forty_2015} included 90 papers in their final set. However, we excluded their mapped paper S86 from our seed set because it is a book chapter, and we did not find evidence of publication in a scientific journal or conference to be approved in IC1 (see Section \ref{sec:smp-studysel}). As suggested in the guidelines, the original study itself \citep{cruz_forty_2015} was included, obtaining a seed set of 90 studies.
    \item \textbf{Use Google Scholar to search for papers and apply forward snowballing, without iteration:} We used the Publish or Perish 7 tool \citep{harzing_publish_2020} to assist this step. The tool has features related to bibliometric analysis, including retrieving citations from publications using Google Scholar. Thus, we conducted the forward snowballing on the seed set using the tool in August 2020 and exported the results for treatment in JabRef\footnote{https://www.jabref.org/}, a bibliographic reference manager. Also, a new forward snowballing step was conducted in January 2021, aiming at covering research from 2020. All screening steps were conducted using JabRef.
    \item \textbf{Include more than one researcher in the initial screening to minimize the risk of removing studies that should be included (false negatives):} One researcher was included to assist in the initial screening of studies and discussions were held with a third researcher.
\end{itemize}

\subsection{Study Selection}
\label{sec:smp-studysel}
\cite{petersen_guidelines_2015} argue that only studies that are relevant to answer the RQs must be considered. Our inclusion criteria (IC1) consists of \textbf{primary studies published in journals, conferences, and workshops reporting SE research using psychometric instruments regarding personality that were published after 2010}\footnote{\cite{cruz_forty_2015} already had mapped 1970 to 2010.}. The exclusion criteria applied to filter the raw set of studies from forward snowballing are presented in Table \ref{IE-table}.

\begin{table}[h] \footnotesize
 \centering
 {\renewcommand\arraystretch{1}
 \caption{Exclusion criteria.}
 \label{IE-table}
 \begin{tabular}{ l l }
 \hline
  \cline{1-1}\cline{2-2}  
    \multicolumn{1}{|p{2cm}|}{\cellcolor[HTML]{EFEFEF}\textbf{Criteria} \centering } &
    \multicolumn{1}{p{9cm}|}{\cellcolor[HTML]{EFEFEF}\textbf{Description} \centering }
  \\  
  \cline{1-1}\cline{2-2}  
    \multicolumn{1}{|p{2cm}|}{EC1} &
    \multicolumn{1}{p{9cm}|}{Papers that are not written in English.}
  \\  
  \cline{1-1}\cline{2-2}  
    \multicolumn{1}{|p{2cm}|}{EC2} &
    \multicolumn{1}{p{9cm}|}{Grey literature. Such as books, theses (bachelor's degree, MSc or PhD), technical reports, occasional papers, and manuscripts without peer-review evidence.}
  \\  
  \cline{1-1}\cline{2-2}  
    \multicolumn{1}{|p{2cm}|}{EC3} &
    \multicolumn{1}{p{9cm}|}{Papers that are only available in the form of abstracts, posters, short versions, and presentations. First, we check whether the paper's information is a short version according to the venue. If not available, we excluded papers with less than six pages.}
  \\ 
  \cline{1-1}\cline{2-2}  
    \multicolumn{1}{|p{2cm}|}{EC4} &
    \multicolumn{1}{p{9cm}|}{Papers that did not include in their title or abstract terms defined by \cite{cruz_forty_2015} as regarding personality. There are: ‘‘personality’’, ‘‘psychological typology’’, ‘‘psychological types’’, ‘‘temperament type’’, and ‘‘traits’’.}
  \\  
  \cline{1-1}\cline{2-2}  
    \multicolumn{1}{|p{2cm}|}{EC5} &
    \multicolumn{1}{p{9cm}|}{Papers addressing other psychometric constructs (\eg behavior, cognition, abilities, roles, etc.), not corresponding to the adopted definition of personality (\textit{cf.} Section \ref{sec:background-personalitySE}).
    }
  \\
   \cline{1-1}\cline{2-2}  
    \multicolumn{1}{|p{2cm}|}{EC6} &
    \multicolumn{1}{p{9cm}|}{Papers that do not meet the inclusion criteria, \ie papers that do not contribute to SE.}
    \\
   \cline{1-1}\cline{2-2}  
    \multicolumn{1}{|p{2cm}|}{EC7} &
    \multicolumn{1}{p{9cm}|}{Papers that use secondary personality data (available datasets, reused data collected in previous studies, etc).}
 \\
 \hline
 \end{tabular} }
\end{table}

This mapping study aims to provide an overview of the use of psychometric instruments in SE research published in peer-reviewed venues. Therefore, we focus on classifying the type of contribution by discovering objectives, the use of psychometric instruments, and the type of research to understand the overall publication landscape without applying a formal quality assessment. The procedure involved reading titles and abstracts and looking for evidence of psychometric instruments. If it was not enough for clarification, the paper’s introduction and the conclusion were read. Still, if not sufficient, the full text of the study was read.

As shown in Table \ref{IE-table}, we only included papers written in English (EC1), peer-reviewed (EC2), and complete (EC3). Due to the volume of papers, it was necessary to adopt an objective and impartial filter strategy based on keywords in the title and abstract (EC4). Focusing on the scope of the study itself, papers that did not deal with personality or related concepts (EC5) in SE studies (EC6) were also eliminated. Papers that used simulated or secondary personality data without applying a psychometric instrument (EC7) were also not considered.

\subsection{Data Extraction and Classification Scheme}
\label{sec:smp-def}

The data extracted from each paper of the final set is shown in Table \ref{DEF}.

\begin{table}[h] \footnotesize
 \centering
 {\renewcommand\arraystretch{1}
 \caption{Data Extraction Form.}
 \label{DEF}
 \begin{tabular}{ l l }
 \hline
  \cline{1-1}\cline{2-2}  
    \multicolumn{1}{|p{3cm}|}{\cellcolor[HTML]{EFEFEF}\textbf{Information} \centering } &
    \multicolumn{1}{p{8cm}|}{\cellcolor[HTML]{EFEFEF}\textbf{Description} \centering }
  \\  
  \cline{1-1}\cline{2-2}  
    \multicolumn{1}{|p{3cm}|}{Study Metadata} &
    \multicolumn{1}{p{8cm}|}{Paper title, author's information, venue, psychometric instrument (name, version, and application process), and year of publication.}
  \\  
  \cline{1-1}\cline{2-2}  
    \multicolumn{1}{|p{3cm}|}{Objective (RQ1a)} &
    \multicolumn{1}{p{8cm}|}{Study objective: we employed open coding \citep{glaser1992} to extract information.}
  \\ 
  \cline{1-1}\cline{2-2}  
    \multicolumn{1}{|p{3cm}|}{Limitations (RQ1b)} &
    \multicolumn{1}{p{8cm}|}{Limitations on the use of psychometric instruments (if exists), such as what were the difficulties of adoption/application and data interpretation. We employed open coding \citep{glaser1992} to extract data.}
  \\
  \cline{1-1}\cline{2-2}  
    \multicolumn{1}{|p{3cm}|}{Purpose of the psychometric instrument in the study (RQ1c)} &
    \multicolumn{1}{p{8cm}|}{What constructs represent the purpose of the psychometric instrument in the study. In SE, constructs are derived from one of the classes: \textit{people}, \textit{organizations}, \textit{technologies}, \textit{activities}, or \textit{software systems} \citep{sjoberg_building_2008}. We employed open coding \citep{glaser1992} to extract data.}
  \\  
  \cline{1-1}\cline{2-2}  
    \multicolumn{1}{|p{3cm}|}{Research Type (RQ1d)} &
    \multicolumn{1}{p{8cm}|}{For research type facets we used the taxonomy proposed by \cite{wieringa_requirements_2005}, containing the following categories: \textit{evaluation research}, \textit{solution proposal}, \textit{validation research}, \textit{philosophical paper}, \textit{opinion paper}, or \textit{experience paper}. \cite{petersen_guidelines_2015} recommendations were followed in this categorization.}
  \\  
  \cline{1-1}\cline{2-2}  
    \multicolumn{1}{|p{3cm}|}{Empirical Evaluation (RQ1e)} &
    \multicolumn{1}{p{8cm}|}{Classification of the empirical study in the following categories of \cite{wohlin_experimentation_2012}: \textit{experiment/quasi-experiment}, \textit{case study}, or \textit{survey}.}
  \\ 
  \hline
 \end{tabular} }
\end{table}

\subsection{Applying the Systematic Mapping Protocol}
\label{sec:applying-smp}

\begin{figure}[h]
\centering
\includegraphics[width=10cm]{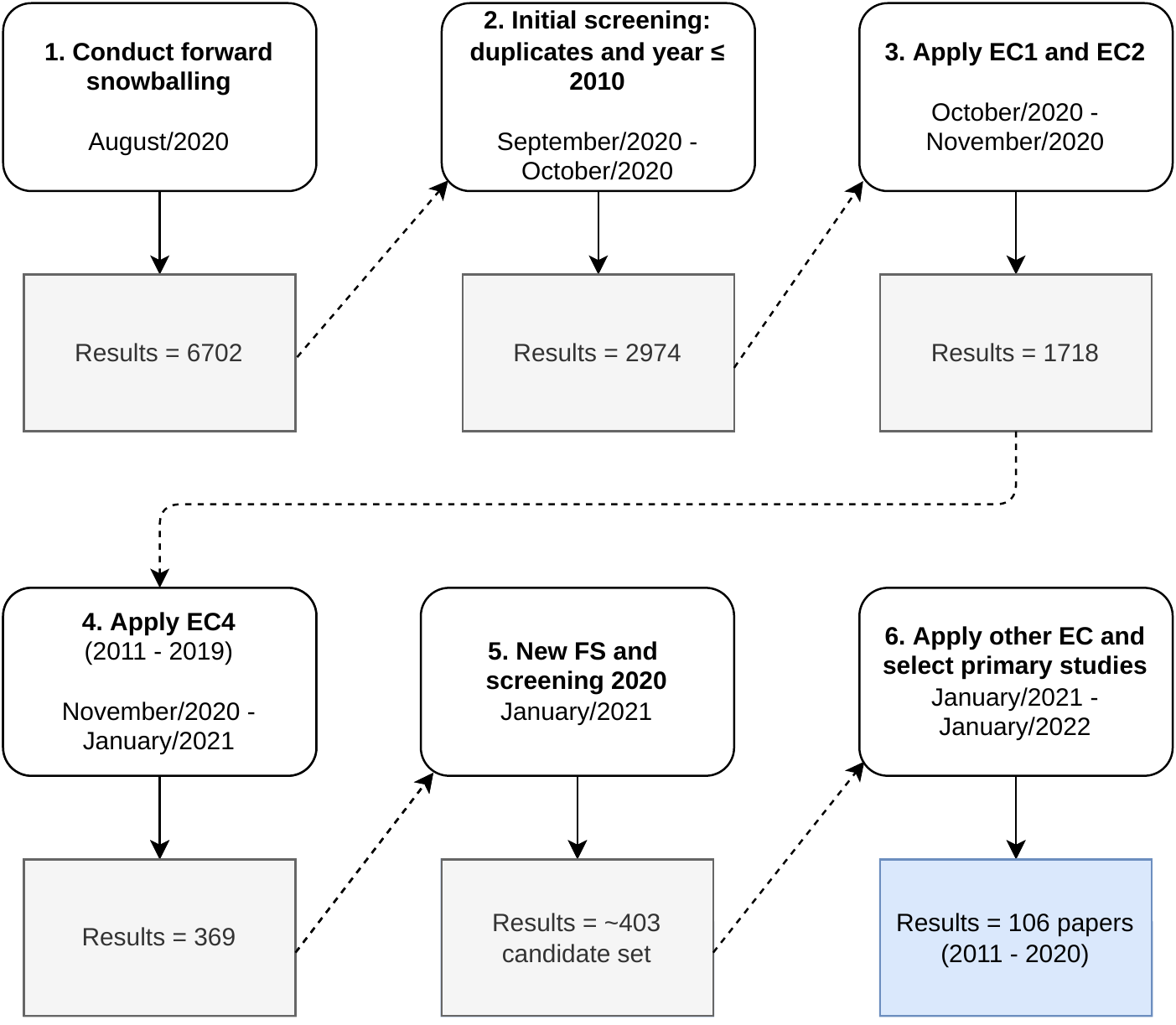}
\caption{Steps of mapping execution.}
\label{fig:current-step}
\end{figure}

The first step to execute the mapping protocol was to conduct forward snowballing on the seed set as described in the Subsection \ref{sec:smp-update}, which generated 6702 entries (step 1 of Figure \ref{fig:current-step}). Between September and October of 2020 we conducted an initial screening of duplicates and of studies with year less than or equal to 2010, given that \cite{cruz_forty_2015} cover a range from 1970 to 2010.

Many entries were provided by Google Scholar/Publish or Perish 7 export feature with incomplete or incorrect data (e.g., journal studies categorized in the entry as books or miscellaneous, and truncated title or abstracts). After removing duplicate entries and when possible, the data for each publicated study were complemented and registered in JabRef to perform a more reliable exclusion per year. The result of this initial screening resulted in 2974 entries (step 2 of Figure \ref{fig:current-step}).

Thereafter, another screening was conducted regarding the exclusion criteria EC1 and EC2. The removal was performed based on the metadata provided in the title, abstract, and journal/booktitle field entries. When it was not possible to easily identify, a verification was made through the URL of the entry or by searching the source on the internet. This exclusion was conducted between October and November of 2020 (step 3 of Figure \ref{fig:current-step}). These exclusion steps reduced the set to 1718 entries.

Thereafter, we removed entries from 2020 and applied EC4, which resulted in 369 entries of 2011 to 2019, step 4 of Figure \ref{fig:current-step}). The removal was conducted aiming at completing our ten-year coverage by replacing them with a new forward snowballing conducted in January 2021 to more consistently include research from 2020. In this new snowballing, we considered only papers from 2020 and applied the same previous ECs, which resulted in a candidate set of 403 entries (step 5 of Figure \ref{fig:current-step}).

In 2021, we applied the other exclusion criteria (EC3, EC5, EC6, and EC7) by reading the remaining studies in the candidate set and extracting data from the selected ones (step 6 of Figure  \ref{fig:current-step}). Two additional researchers assisted in this step covering two years each (2017 to 2018 and 2019 to 2020), using a prepared web form with detailed advice for data extraction. All exclusions/inclusions and extracted data were carefully reviewed. As a result, 106 studies were included and had their data extracted. The list of papers and the extracted data can be found packaged online in \cite{felipe_psychometric_2022}.

In the next section we present the results extracted and organized by our RQs. Bibliographic references of the selected studies are presented in \ref{appendix:final-set}.



\section{Systematic Mapping Study Results}
\label{sec:Results}

This section presents the results of the mapping study. First, we provide an overview of the included studies, followed by answers of defined RQs based on the information extracted from the included studies (Subsections \ref{sec:results-overview} to \ref{sec:results-empirical}). 

\subsection{Overview}
\label{sec:results-overview}

\begin{figure}[h]
\centering
\includegraphics[width=10cm]{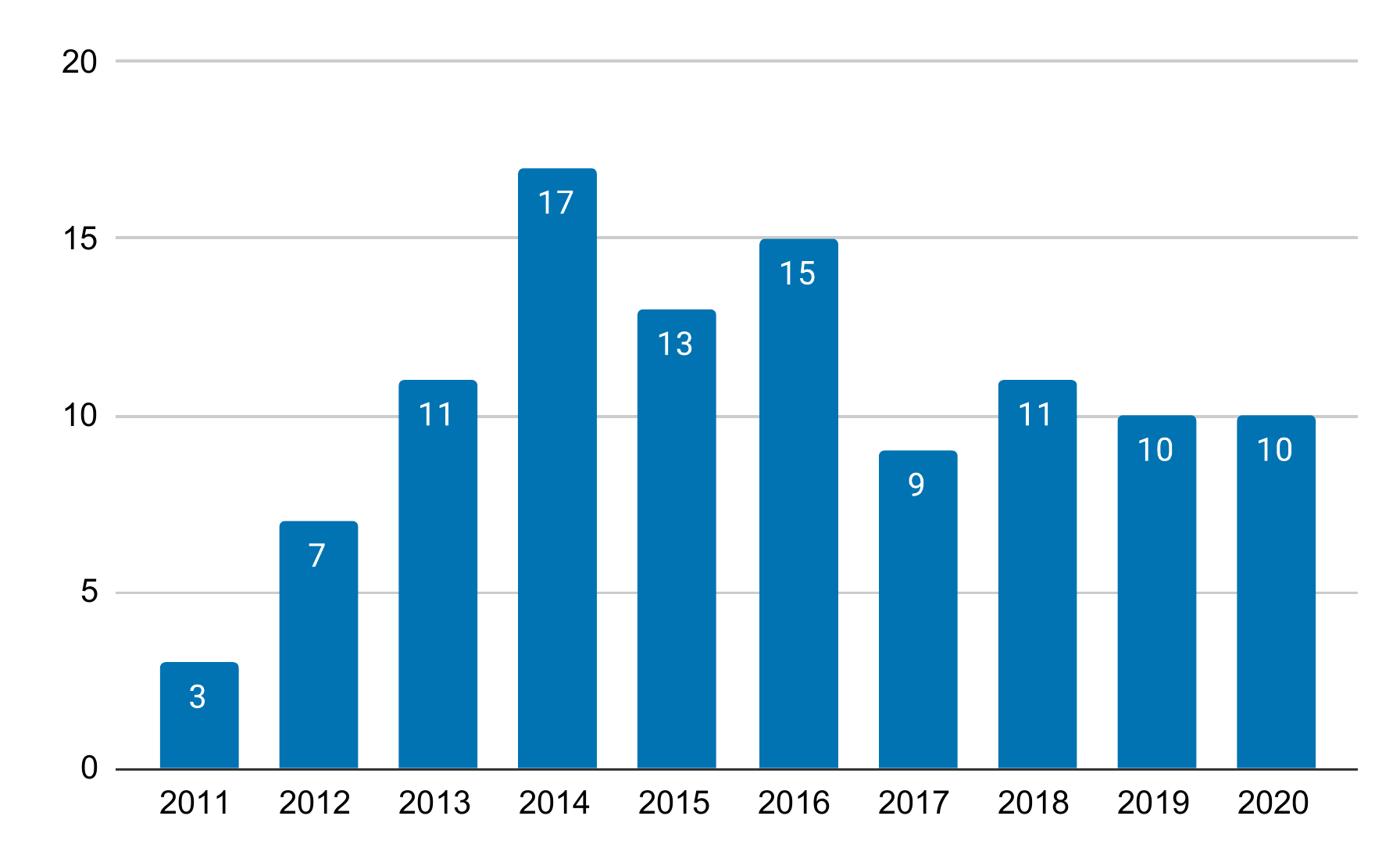}
\caption{Temporal distribution of selected primary studies.}
\label{fig:temporal-distribution}
\end{figure}

We identified 106 additional primary studies that employ psychometric instruments in SE research on personality, ranging from 2011 to 2020\footnote{We refer to the identification of studies from our protocol as S(NUMBER), where (NUMBER) begins to account from 91, given that \cite{cruz_forty_2015} has 90 mapped studies.}. The temporal distribution of the studies is depicted in Figure \ref{fig:temporal-distribution}. We can observe that the time range of 2014 to 2016 holds the highest frequency of studies. Indeed, when screening and extracting data process, this period required more effort due to also having a larger volume of papers to be analyzed. 

Nevertheless, it is possible to observe that we identified 40 additional papers ranging from 2017 to 2020. In fact, while we found 106 studies in our investigated ten-year range (2011 to 2020), \cite{cruz_forty_2015} found 90 studies in the previous 40-year range (1970 to 2010). 

Most of the studies (62 out of 106) were published in journals, which reinforces our effort in data extraction due to fewer restrictions on document size generally required. Followed by conference (39 out of 106) and workshop (5 out of 106) publications. The latter could have a low number due to more restrictions in document size, hence not approved by our ECs, as we were looking for complete research papers.

\subsection{RQ1. \rqmap}

\begin{figure}[h]
\centering
\includegraphics[width=10cm]{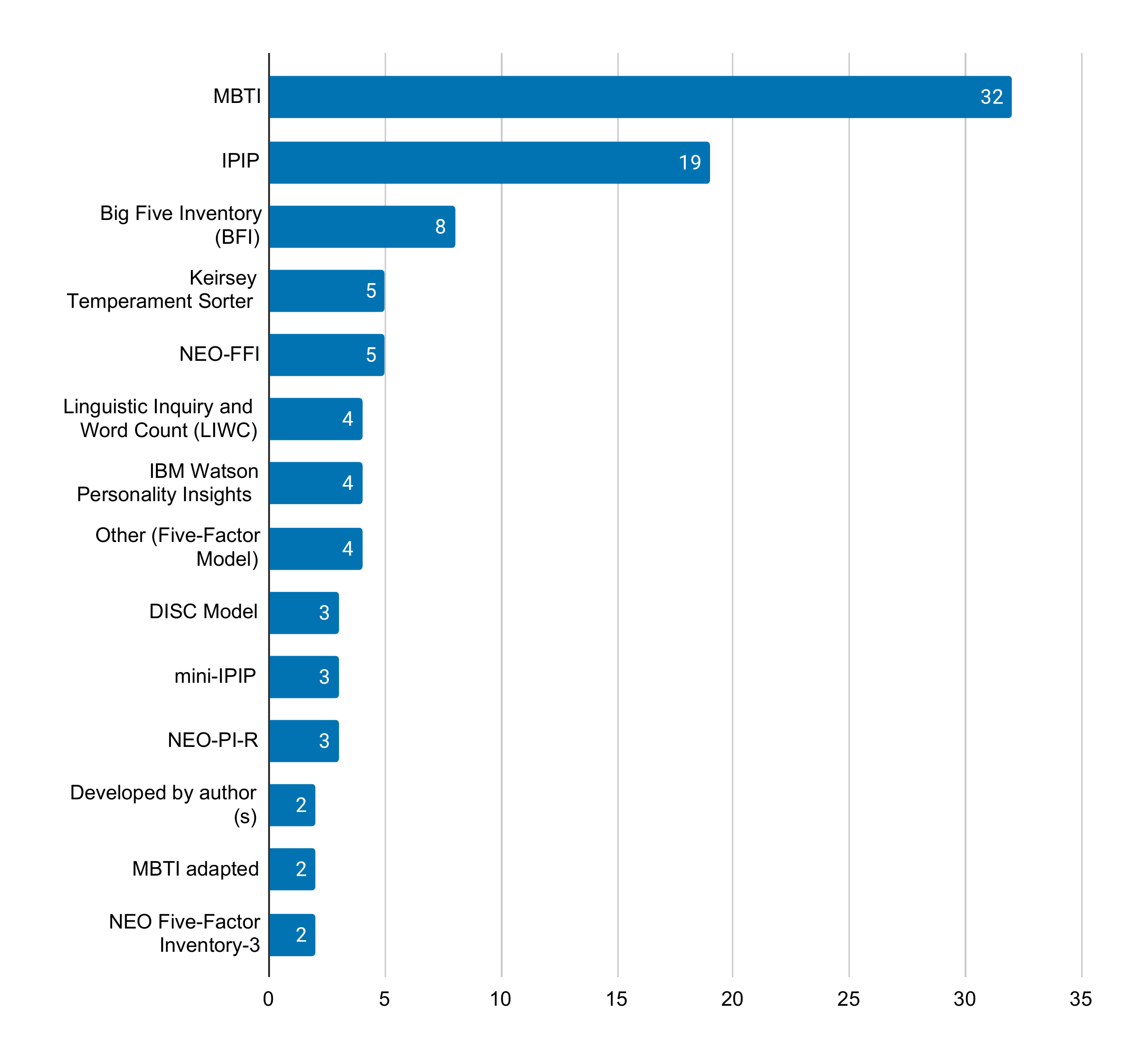}
\caption{Frequency of used psychometric instruments.}
\label{fig:frequency-pis}
\end{figure}

The frequency of instruments is illustrated in Figure \ref{fig:frequency-pis}. The most used instrument refers to versions of the Myers-Briggs Type Indicator (MBTI), a finding also reported in \cite{cruz_forty_2015}. It is noteworthy that we observed a recent reduction regarding the use of MBTI, which we will further address in Section \ref{sec:results-discussion}.  For illustration purposes, instruments used in only one study are not shown in the chart. An overview of the complete list can be found in Table \ref{tab:pis}.

To facilitate the instruments' categorization, we compared details of the versions reported through bibliographic references and specific information available to understand if they were referring to the same instrument. In these cases, we kept the details on the version in our repository but consolidated them for analysis purposes (\eg MBTI, mapped across multiple versions). 

It is also noteworthy that a variety of instruments that operationalize the Five-Factor Model of personality are used (\eg IPIP, BFI, NEO-FFI, mini-IPIP, NEO-PI-R, and NEO Five-Factor Inventory-3). Together with the MBTI, this model stand out as the main theoretical background for instruments applied in SE research. Although the instruments based on the Five-Factor Model are the majority in sum, we consider each one of the instruments shown in Table \ref{tab:pis} as an individual instrument in our analysis.

\begin{table}[h!]
\centering
\caption{Psychometric instruments mapped organized by studies.}
\label{tab:pis}
\resizebox{\textwidth}{!}{%
\begin{tabular}{|l|l|r|}
\hline
\rowcolor[HTML]{EFEFEF} 
\multicolumn{1}{|c|}{\cellcolor[HTML]{EFEFEF}\textbf{Pyschometric instrument}} & \multicolumn{1}{c|}{\cellcolor[HTML]{EFEFEF}\textbf{Studies}} & \multicolumn{1}{c|}{\cellcolor[HTML]{EFEFEF}\textbf{Count}} \\ \hline
MBTI & \begin{tabular}[c]{@{}l@{}}S91, S93, S99, S101, S110, S111, S112, S119, S120, \\ S122, S123, S126, S127, S129, S130, S137, S138,\\  S151, S155, S158, S162, S163, S165, S167, S168,\\  S169, S172, S173, S177, S179, S181, S196\end{tabular} & 32 \\ \hline
IPIP & \begin{tabular}[c]{@{}l@{}}S92, S106, S113, S124, S128, S132, S135, S136,\\ S139, S146, S150, S161, S164, S171, S174, S178,\\  S182, S186, S195\end{tabular} & 19 \\ \hline
Big Five Inventory (BFI) & S100, S104, S114, S125, S141, S175, S183, S189 & 8 \\ \hline
Keirsey Temperament Sorter (KTS) & S96, S103, S108, S131, S147 & 5 \\ \hline
NEO-FFI & S117, S133, S140, S153, S174 & 5 \\ \hline
Linguistic Inquiry and Word Count (LIWC) & S116, S134, S149, S194 & 4 \\ \hline
IBM Watson Personality Insights & S152, S176, S180, S188 & 4 \\ \hline
Other (Five-Factor Model) & S107, S143, S145, S192 & 4 \\ \hline
DISC Model & S109, S148, S157 & 3 \\ \hline
mini-IPIP & S118, S142, S156 & 3 \\ \hline
NEO-PI-R & S95, S102, S110 & 3 \\ \hline
Developed by author(s) & S121, S184 & 2 \\ \hline
MBTI adapted & S94, S154 & 2 \\ \hline
NEO Five-Factor Inventory-3 & S97, S98 & 2 \\ \hline
NEO-PI3 & S105 & 1 \\ \hline
IPIP based & S115 & 1 \\ \hline
Five-Factor Stress & S144 & 1 \\ \hline
Other (Unspecified) & S159 & 1 \\ \hline
Student Styles Questionnaire & S166 & 1 \\ \hline
EPQ-R & S170 & 1 \\ \hline
IPI & S185 & 1 \\ \hline
16 Personality Factors (16PF) & S187 & 1 \\ \hline
HEXACO Model & S191 & 1 \\ \hline
\cellcolor[HTML]{FFFFFF}Quick Big Five (QBF) & S193 & 1 \\ \hline
\end{tabular}%
}
\end{table}

\subsection{RQ2. \rqmapA}
During the data extraction, it was possible to observe the following major objectives with open and axial coding procedures \citep{glaser1992}\footnote{An overview of qualitative coding, in the context of grounded theory for SE research, has been offered by \cite{stol_grounded_2016}}. 

\begin{figure}[H]
\centering
\includegraphics[width=9cm]{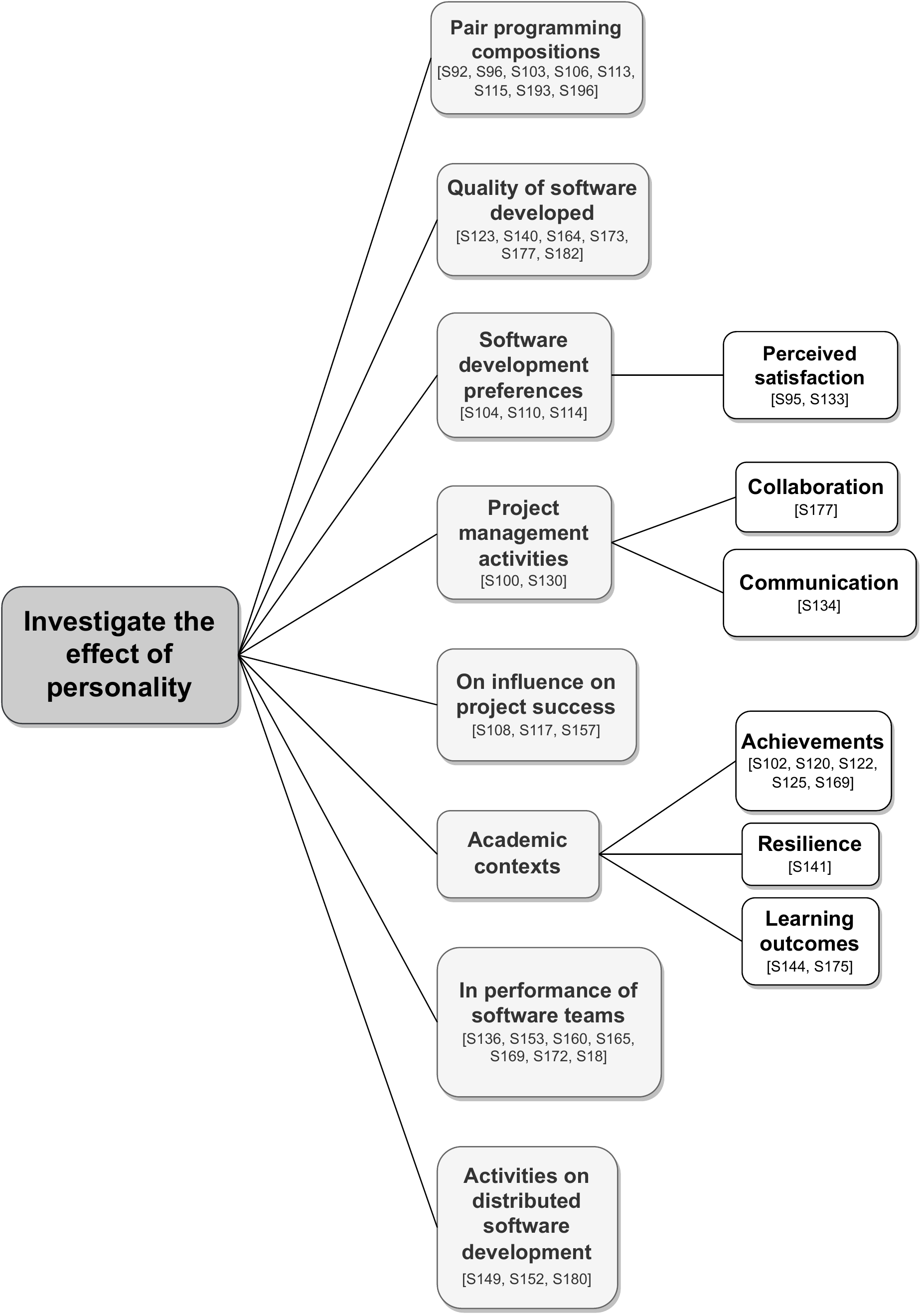}
\caption{Tree structure of \textit{Investigate the effect of personality} major code.}
\label{fig:rq1a-1}
\end{figure}

\textit{Investigate the effect of personality}: in this major coded objective, data on personality is used as an intervention to investigate phenomena. We identified this one with several specificities illustrated, as shown in Figure \ref{fig:rq1a-1} and listed below for minor objectives with more than two papers. Such specificities aim to \textit{investigate the effect of personality}:
\begin{itemize}
    \item in pair programming teams compositions [S92, S96, S103, S106, S113, S115, S193, S196]; 
    \item in the quality of software developed [S95, S123, S133, S140, S164, S173, S177, S182] and their respective perceived satisfaction [S95, S133];
    \item regarding influence on project management activities [S100, S130], including communication [S134] and collaboration [S177];
    \item in academic contexts, such as analyzing achievements [S102, S120, S122, S125, S169], resilience [S141] or learning outcomes [S144, S175] of students;
    \item in development preferences [S104, S110, S114];
    \item on the influence on project success [S108, S117, S157];
    \item in performance of software teams [S136, S153, S160, S165, S169, S172, S187]; and
    \item in activities on distributed software development [S149, S152, S180].
\end{itemize}

Furthermore, the effect of personality has also been investigated in other topics, such as the use of CASE [S101, S170] and static analysis tools [S143], implementing a new technology [S109], performance of software engineer [S127, S146], programming styles [S135, S150], software testing performance [S105], team climate and productivity [S139, S195], requirements engineering activities [S142], knowledge management activities [S159, S162], software engineer burnout tendencies [S179], collaboration [S176], creativity [S189], and on the use of software repositories [S194].

\begin{figure}[h]
\centering
\includegraphics[width=12cm]{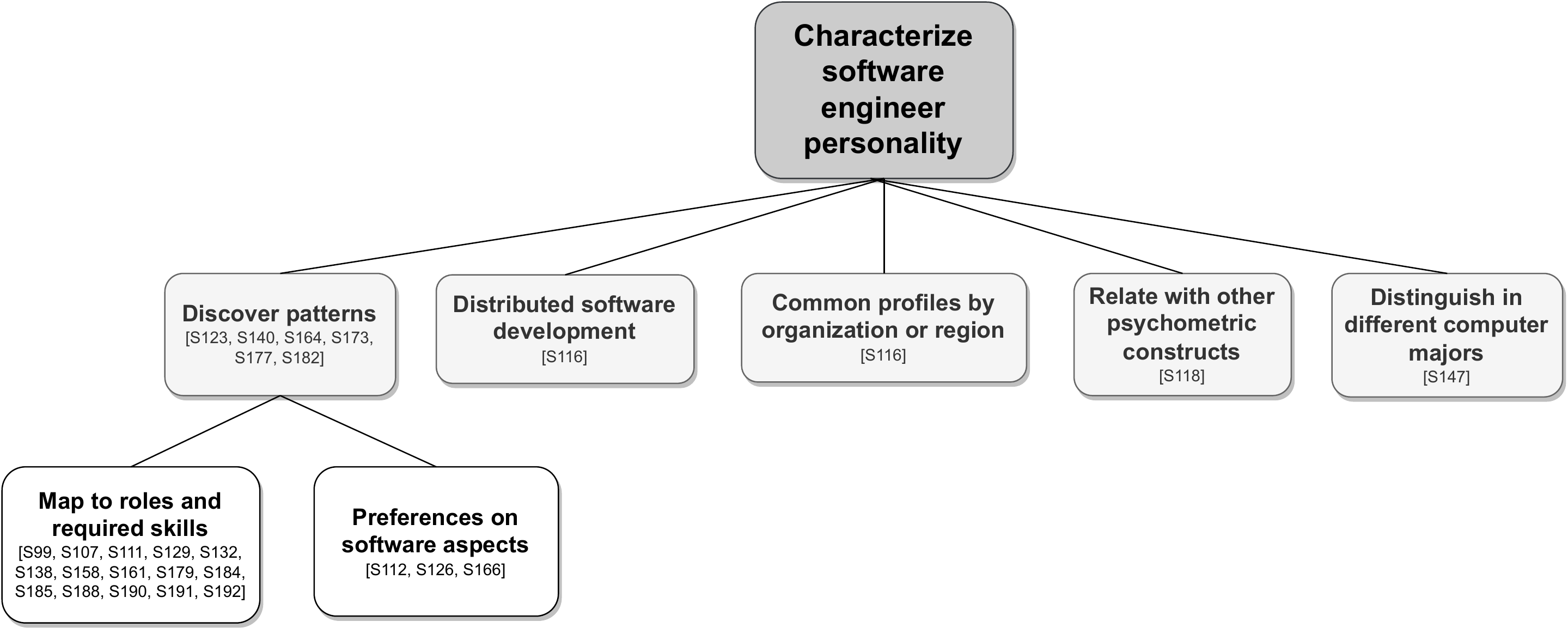}
\caption{Tree structure of \textit{characterize software engineer personality} major code.}
\label{fig:rq1a-2}
\end{figure}

\textit{Characterize software engineer personality}:
as shown in Figure \ref{fig:rq1a-2}, some studies aimed to discover patterns in personality of software engineering  professionals [S93, S119] and more specifically to mapping these patterns to roles and required skills [S99, S107, S111, S129, S132, S138, S158, S161, S179, S184, S185, S188, S190, S191, S192] or preferences in software aspects [S112, S126, S166]. Characteristics of software engineering personality are also explored in distributed software development [S116]; considering common profiles of personality by region or organization-context [S131, S137]; to relate them with other psychometric constructs over time [S118]; and with respect to distinguishing profiles within different computer major courses [S147].

\begin{figure}[h]
\centering
\includegraphics[width=4cm]{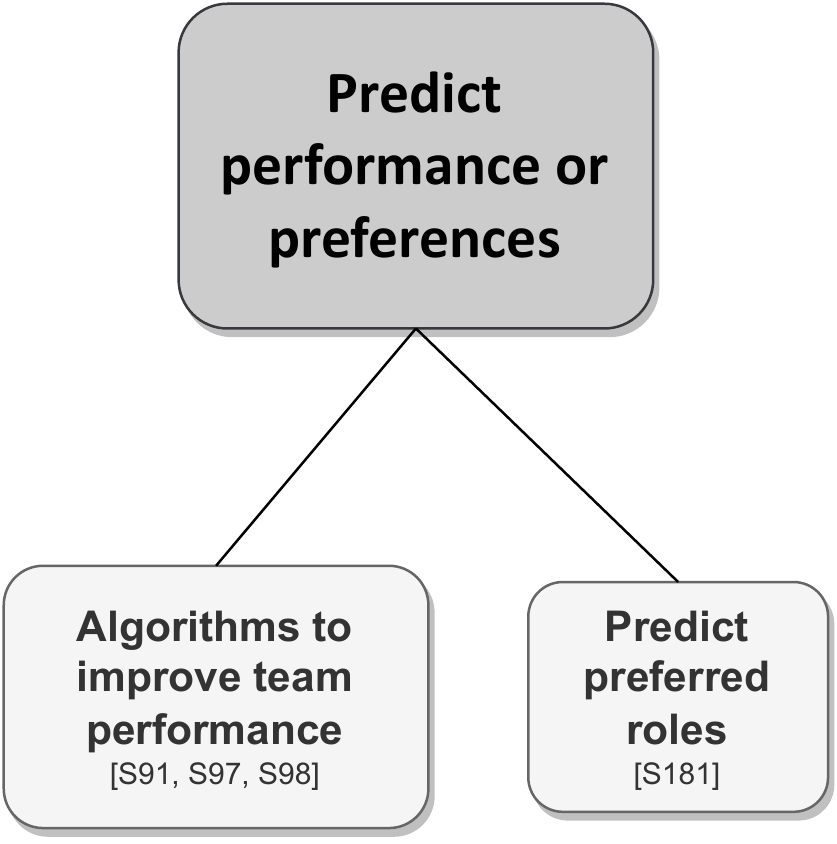}
\caption{Tree structure of \textit{predict performance or preferences} major code.}
\label{fig:rq1a-5}
\end{figure}

\textit{Predict performance or preferences}: four studies used information on personality to perform predictions (see Figure \ref{fig:rq1a-5}). This objective involves development team's performance algorithms and tools in order to optimize resources [S91, S97, S98] and predict preferred roles of software engineering professionals [S181].

\begin{figure}[h]
\centering
\includegraphics[width=12cm]{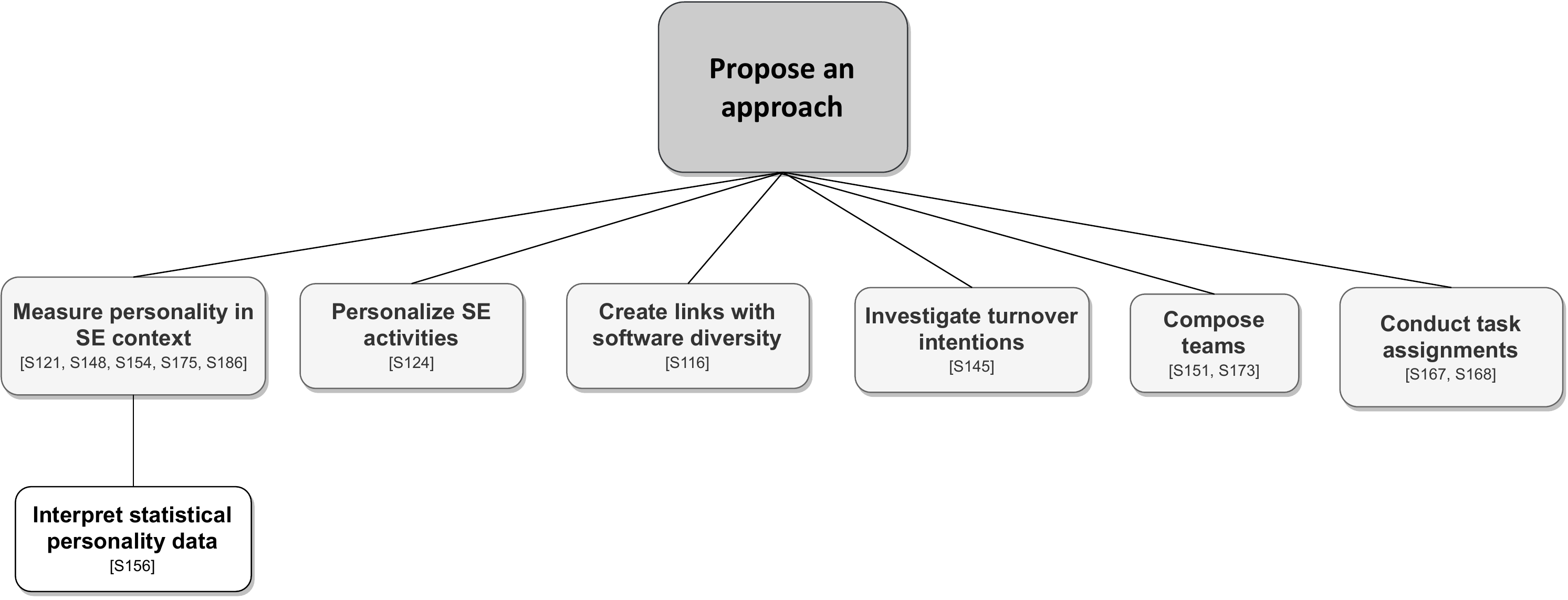}
\caption{Tree structure of \textit{propose an approach} major code.}
\label{fig:rq1a-3}
\end{figure}

\textit{Propose an approach}: in this objective, some proposal is given to support research or practical activities in SE (Figure \ref{fig:rq1a-3}). For instance, proposing approaches to measure personality [S121, S148, S154, S175, S186] and to interpret statistically data in domain of SE [S156]; to personalize SE activities [S124]; to create links between human factors (like personality) with software diversity [S128]; to investigate turnover intentions [S145]; to relate individual characteristics to development performance [S183]; to perform team compositions [S151, S173]; and to conduct task assignments [S167, S168].

\begin{figure}[]
\centering
\includegraphics[width=12cm]{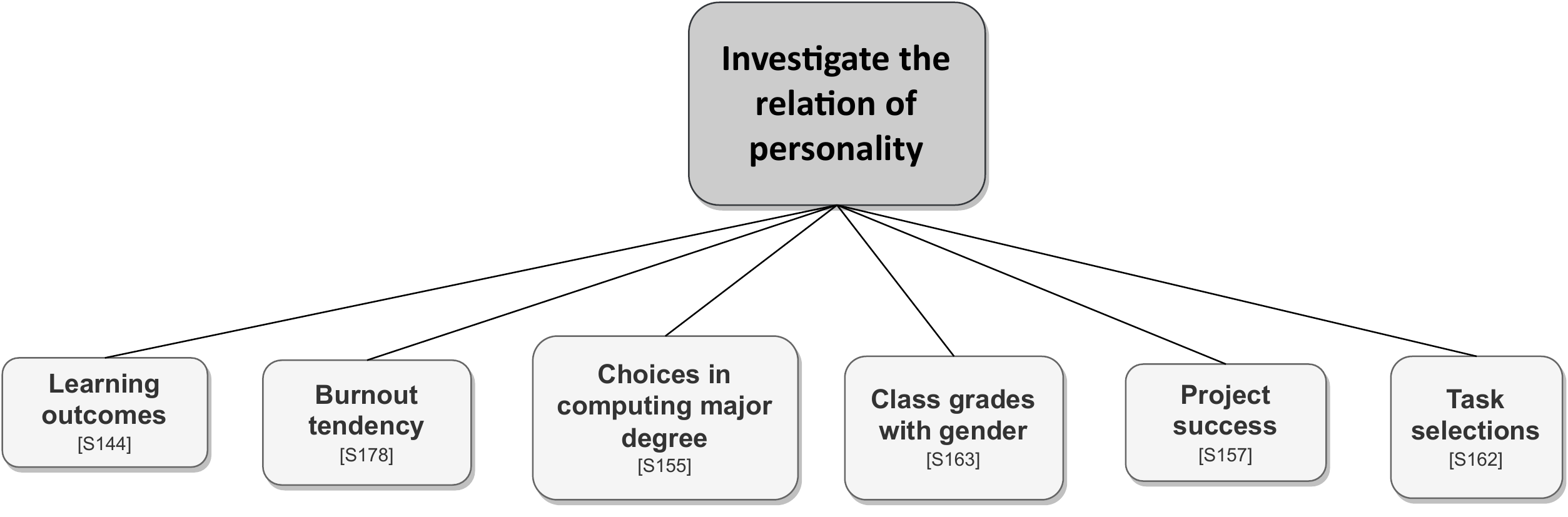}
\caption{Tree structure of \textit{investigate the relationship of personality} major code.}
\label{fig:rq1a-4}
\end{figure}

\textit{Investigate the relation of personality}: this objective is related to check whether personality has any relationship with a research object (see the structure in Figure \ref{fig:rq1a-4}). Personality relationships are investigated with learning outcomes [S144], burnout tendency [S178], choices in major degree in computing [S155], class grades considering gender [S163], learning effectiveness [S174], project success [S157], and task selections [S162].

It is noteworthy that we decided on a completely new coding without a direct influence from previous results to capture the essence of the new results more precisely \citep{seaman_qualitative_1999}. However, we can compare our mapped objectives with the most frequent research topics mapped by \cite{cruz_forty_2015}. Listing them (highlighted): \textbf{\textit{pair programming}}, in \textit{investigate the effect of personality}; \textbf{\textit{education}}, during the data extraction, we identified studies that contributed to educational purposes or used an academic scenario to reach the research goal in all major codes described earlier; \textbf{\textit{team effectiveness}}, in \textit{predicting team performance}; \textbf{\textit{software process allocation}}, \textbf{\textit{software engineer personality characteristics}}, \textbf{\textit{individual performance}}, and \textbf{\textit{team process}} in all of our major codes. It can show us that these topics are still being researched.

\subsection{RQ3. \rqmapB}

Less than half of the primary studies (43 out of 106) reported limitations related to the adoption of psychometric instruments in SE research. An overview of our coded limitations is described in the following.

\textit{Possible misuse of psychometric instrument:}
the authors of [S95, S132, S152, S157, S187] declare that adopting the psychometric instrument to the objectives of research could affect the validity of the study, but this limitation is mitigated by relying on the literature. Other threats concern not involving psychologists in the research design [S130, S148, S187], use of non-intuitive platforms, and poor instructions on instruments' application [S178]. The lack of a data set for performing benchmarks is also reported [S180].

\textit{Choice of a short version of psychometric instrument:} the statistical power of personality data in the studies may be compromised by the adoption of shorter versions of the instruments, so they could result in less accurate results [S118, S156, S191], hence compromising the research goal. Furthermore,~\cite{graziotin2022psychometrics} reports that shortening validated measurement instruments in a non-systematic way may drastically reduce psychometric reliability and validity properties, to the point that we can not be confident anymore of our interpretation of results.

\textit{Personality may not be a representative construct:} the use of personality as an investigated human factor may not be a good choice for the research design [S128] and encountered correlations may not assure causality [S135, S150].

\textit{Bias on subjects responses:} the authors indicate that subjects’ administration of psychometric instruments can become a threat in cases of factors like lack of honesty or loss of concentration of the subjects [S96, S97, S110, S113, S143, S146, S149, S164, S172, S177, S178, S184, S190, S195] sometimes caused by the absence of control of researcher employing certain empirical approaches (like surveys). This limitation is generally mitigated by ensuring they made aware that the response data obtained is anonymized and used only for research purposes.

\textit{Statistical power of psychometric instrument:} the dichotomous approach of MBTI-based instruments could affect results in scenarios when a person is in a center of a scale due to the instruments' statistical structure [S96, S103, S182]. Others report that personality traits data should be measured by adopting other psychometric instruments in order to achieve better results [S100, S104, S111, S158, S163, S173] but without suggesting any other instrument; in case of identification of personality traits from textual analysis, the amount of chunk text may not be sufficient [S176]. Moreover, the one scale/trait of the psychometric instrument showed low internal validity, being excluded from the study [S170].

\textit{Paid subjects:} participants were paid to participate in the study, which may have influenced them somehow [S105].

\textit{Issues with dictionary to measure personality from text:} adequacy of language dictionaries to measure personality from textual analysis may be a threat. Also, the data extracted to measure personality may not be representative of a person [S116, S134, S194].

\textit{Construction issues on proposed psychometric instrument:} the adequacy of psychometric instrument to SE context may deal with validity issues. These issues were mitigate using expert judgments and a series of incremental refinements [S160].

\subsection{RQ4. \rqmapC}

Regarding SE constructs, we used the framework to describe theories proposed by \cite{sjoberg_building_2008} and its archetypal classes as support for open coding constructs. This framework is largely used in SE research to present theories. In it, the archetypal classes interact together in the following way: an \textit{actor} applies a \textit{technology} (we believe that \textit{intervention} is more appropriate in the context of psychometric instruments) to perform certain \textit{activities} on a \textit{software system}. In Tables \ref{tab:app-actor} to \ref{tab:app-ss} (\ref{appendix:coded-constructs}) we list and describe the coded SE constructs to which the psychometric instruments are related within the identified studies, organized by archetypal class.

\begin{landscape}
\begin{figure}[]
\centering
\includegraphics[width=20cm]{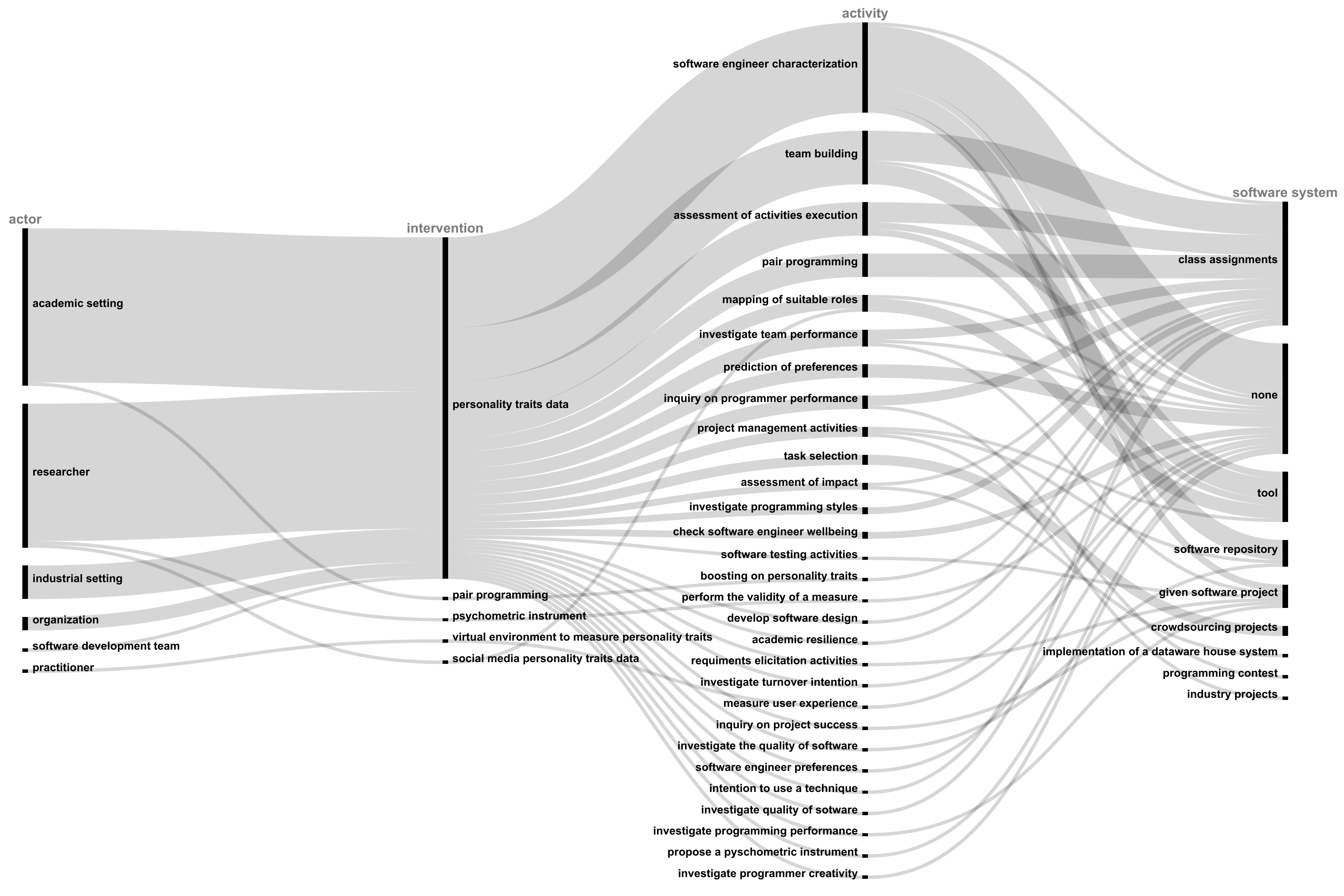}
\caption{SE constructs related to the psychometric instruments}
\label{fig:alluvial-constructs}
\end{figure}
\end{landscape}

Figure \ref{fig:alluvial-constructs} depicts the relationship network between the constructs described in the aforementioned tables. We can observe that constructs of class \textbf{Actor} \textit{researcher} and \textit{academic setting} are the majority, indicating the application of studies in an academic setting due to scope limitations of the research design or for purely investigative purposes by researchers. In the \textbf{Intervention} class, \textit{personality traits data} represents the most common coded construct and is typically measured by some psychometric instrument and used to investigate its influence on SE activities.

Regarding the \textbf{Activity} class, the constructs \textit{software engineer characterization}, \textit{team building}, \textit{assessment of activities execution}, \textit{pair programming}, and \textit{mapping of suitable roles} have a higher frequency. All of them are related to the previously mentioned constructs in classes \textbf{Actor} (\textit{academic setting} and \textit{researcher}) and \textbf{Intervention} (\textit{personality traits data}).

Still, the activities described earlier are also strongly related to the constructs of the \textbf{Software System} class. It is possible to observe that \textit{software engineer characterization}
is typically not related to any specific software system (code \textit{none}), indicating no direct reference or use to software systems in these studies. The code \textit{tool} indicates the use of some technique using software/algorithms/games/logic rules to support the studies. Moreover, \textit{class assignments} were specially related to \textit{team building} and \textit{pair programming}, where SE teams were built based in academic contexts based on personality data. It is noteworthy to mention the use of a \textit{software repository}, in which software artifacts of different kinds are stored.

Please note that RQ1c aims at answering what parts of SE theory the psychometric instruments are related to. There may be similarities with the overall objectives of the mapped studies (RQ1a), but RQ1c is specifically focused on the context of used instruments in the primary studies and their relations to SE theory elements. 

\subsection{RQ5. \rqmapD}

We used the following research type facets taxonomy defined by \cite{wieringa_requirements_2005} and the advice for distinguishing between the categories contained in \cite{petersen_guidelines_2015}.

\begin{itemize}
    \item \textbf{Validation Research}: these papers investigate the properties of a solution proposal that has not yet been implemented in practice.
    \item \textbf{Evaluation Research}: these papers investigate a problem or an implementation of a technique in practice.
    \item \textbf{Solution Proposal}: in these papers, a solution for a problem is proposed, it can be either novel or a significant extension of an existing technique. 
    \item \textbf{Philosophical papers}: these papers sketch a new way of looking at things, a new conceptual framework, etc.
    \item \textbf{Opinion papers}: these papers contain the author’s opinion about what is wrong or good about something,
    how we should do something, etc.
    \item \textbf{Personal experience papers}: in these papers, the emphasis is on what and not on why. The experience may concern
    one project or more, but it must be the author’s personal
    experience.
\end{itemize}

\begin{figure}[h!]
\centering
\includegraphics[width=12cm]{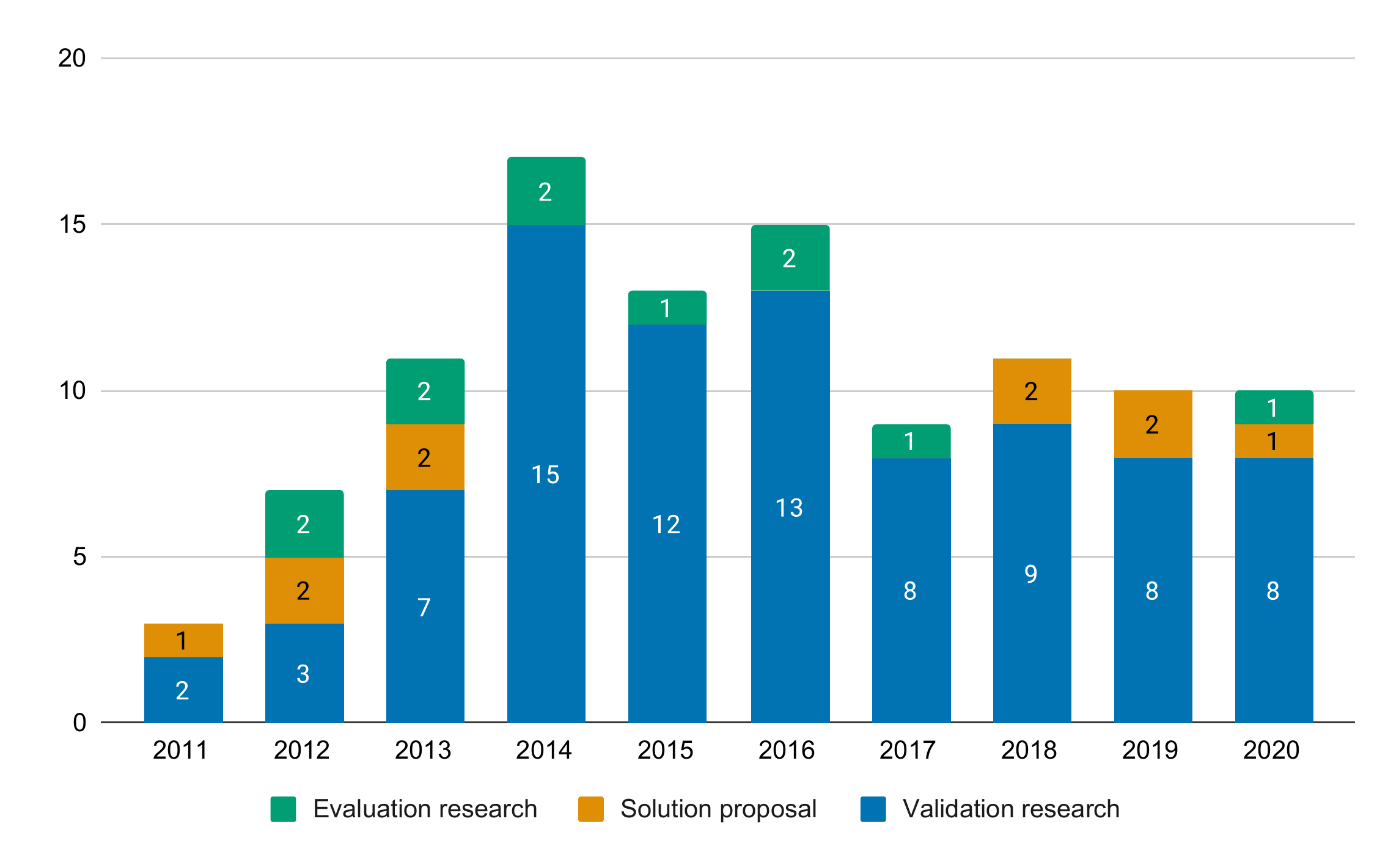}
\caption{Frequency of research type per year.}
\label{fig:research-type}
\end{figure}

Figure \ref{fig:research-type} depicts the distribution of research types facets by year. It is possible to note that validation research over-represented the set of mapped studies either in distribution per year and in total (85 out of 106). This facet includes empirical studies, as well as the less frequent evaluation research (11 out of 106).  The difference between them indicates that most empirical research has been conducted in academic scenarios for initial validation purposes and does not propose and measure new proposals in industrial scenarios. This is somehow expected due to possible difficulties in using industrial practitioner subjects as part of research designs.

Few solution proposals have also been mapped (10 out of 106), which typically represent new research proposals with some limited evaluation required for publication or no presentation of empirical evaluation, therefore they were not classified as either evaluation or validation research. 

\subsection{RQ1e. \rqmapE}
\label{sec:results-empirical}

\begin{figure}[]
\centering
\includegraphics[width=12cm]{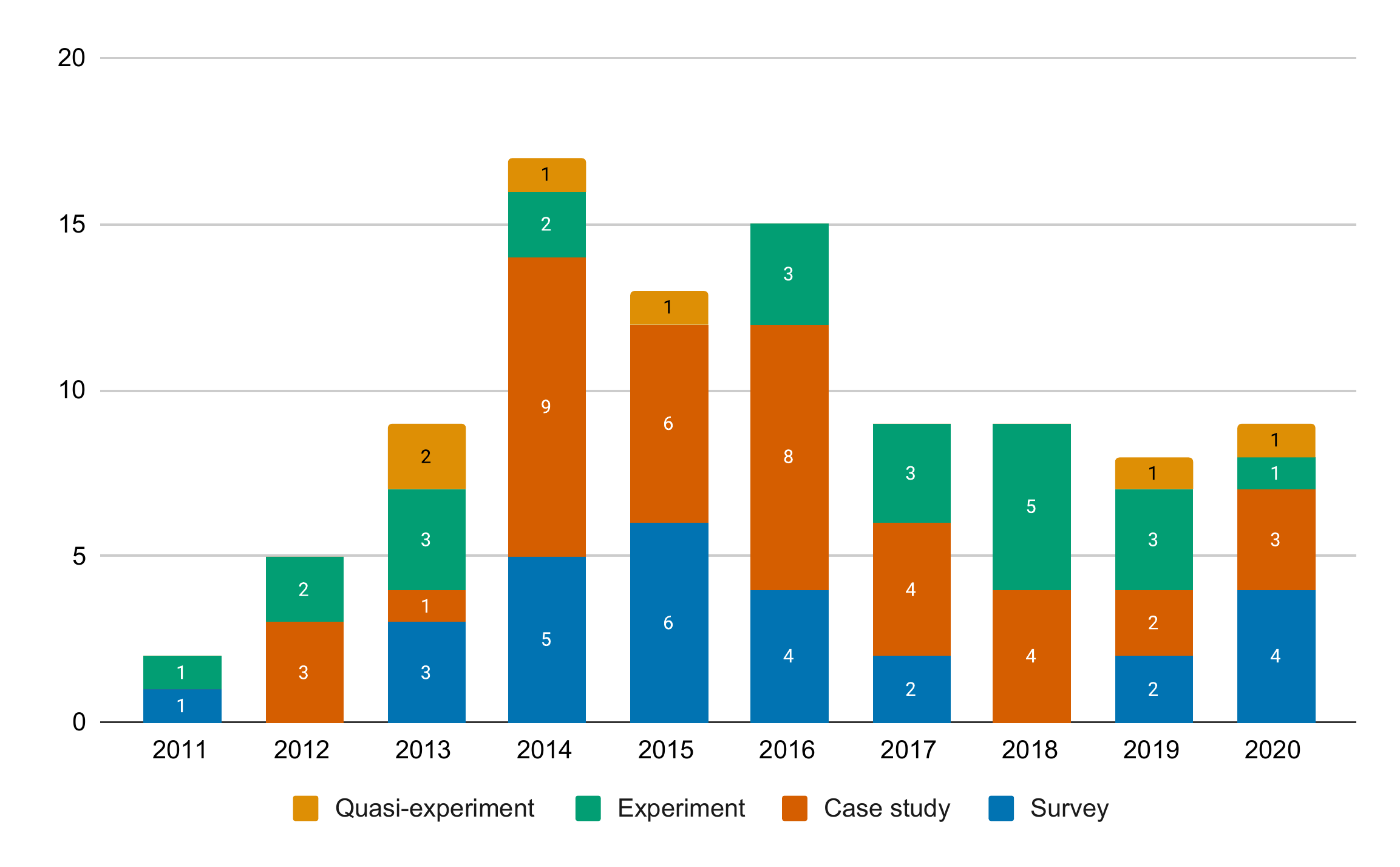}
\caption{Frequency of empirical evaluation types per year.}
\label{fig:empirical-approach}
\end{figure}

Figure \ref{fig:empirical-approach} depicts the frequency of empirical evaluations adopted by year, with 96 out of 106 studies. When analyzing the figure, it is possible to observe that survey and case study strategies have been frequently adopted through the years. In the case of surveys, studies generally use this empirical strategy to apply psychometric instruments.

It is noteworthy that case studies represent a high frequency of empirical approaches. Many studies document the research design as experiments (which we see in figure less frequent), but in an inconsistent way with the definition of experimental/quasi-experimental design \citep{wohlin_experimentation_2012}. In these cases, we classified the empirical evaluation type as case studies. Experiments and quasi-experiments are shown less frequently. This may be explained due to the complexity of handling personality as a variable in controlled experiments.

\section{Discussion}
\label{sec:results-discussion}

To outline how the use of psychometric instruments in SE research on personality evolved over the last fifty years, we manually integrated the data from our study with the corresponding data from the study by \cite{cruz_forty_2015}. Unfortunately, there was no public access to the data extracted in their mapping. However, the paper contains a table listing the psychometric instruments used in each of the papers included in their study. Therefrom, we prepared a new spreadsheet for this particular analysis, which is available in our online repository. For aggregation purposes, similarly to what had been done by \cite{cruz_forty_2015}, we grouped a variety of instruments (\eg IPIP, BFI, and NEO-FFI) that operationalize the Five-Factor Model into the BF/FFM category.

\begin{figure}[h!]
\centering
\includegraphics[width=12cm]{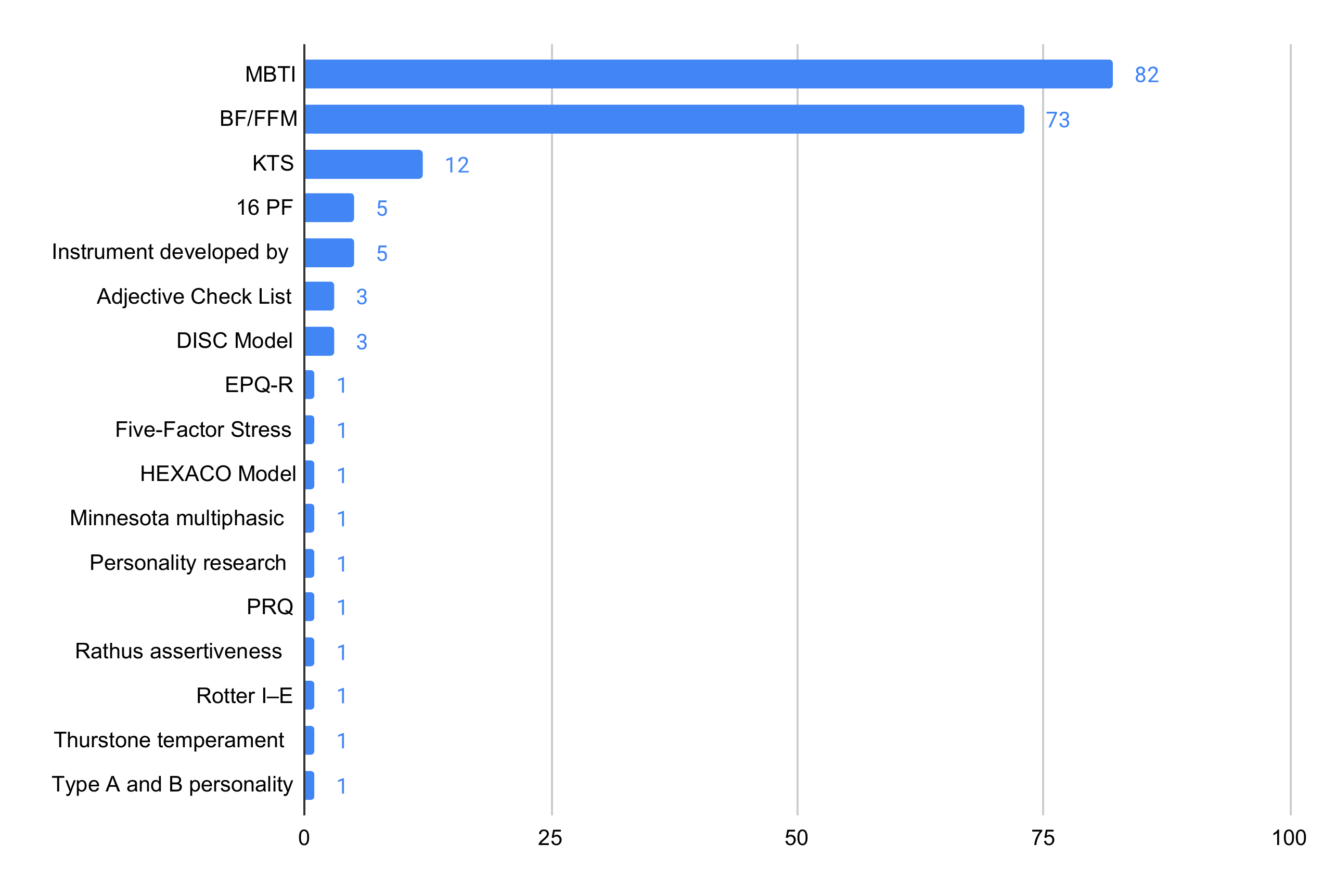}
\caption{Frequency of psychometric instruments adoption.}
\label{fig:all-fifty-freq}
\end{figure}

Figure \ref{fig:all-fifty-freq} shows the overall usage of psychometric instruments in SE research from 1970 to 2020. It is possible to observe that MBTI and FFM stand out as the most used theoretical backgrounds for instruments applied in SE research.

\begin{figure}[h!]
\centering
\includegraphics[width=14cm]{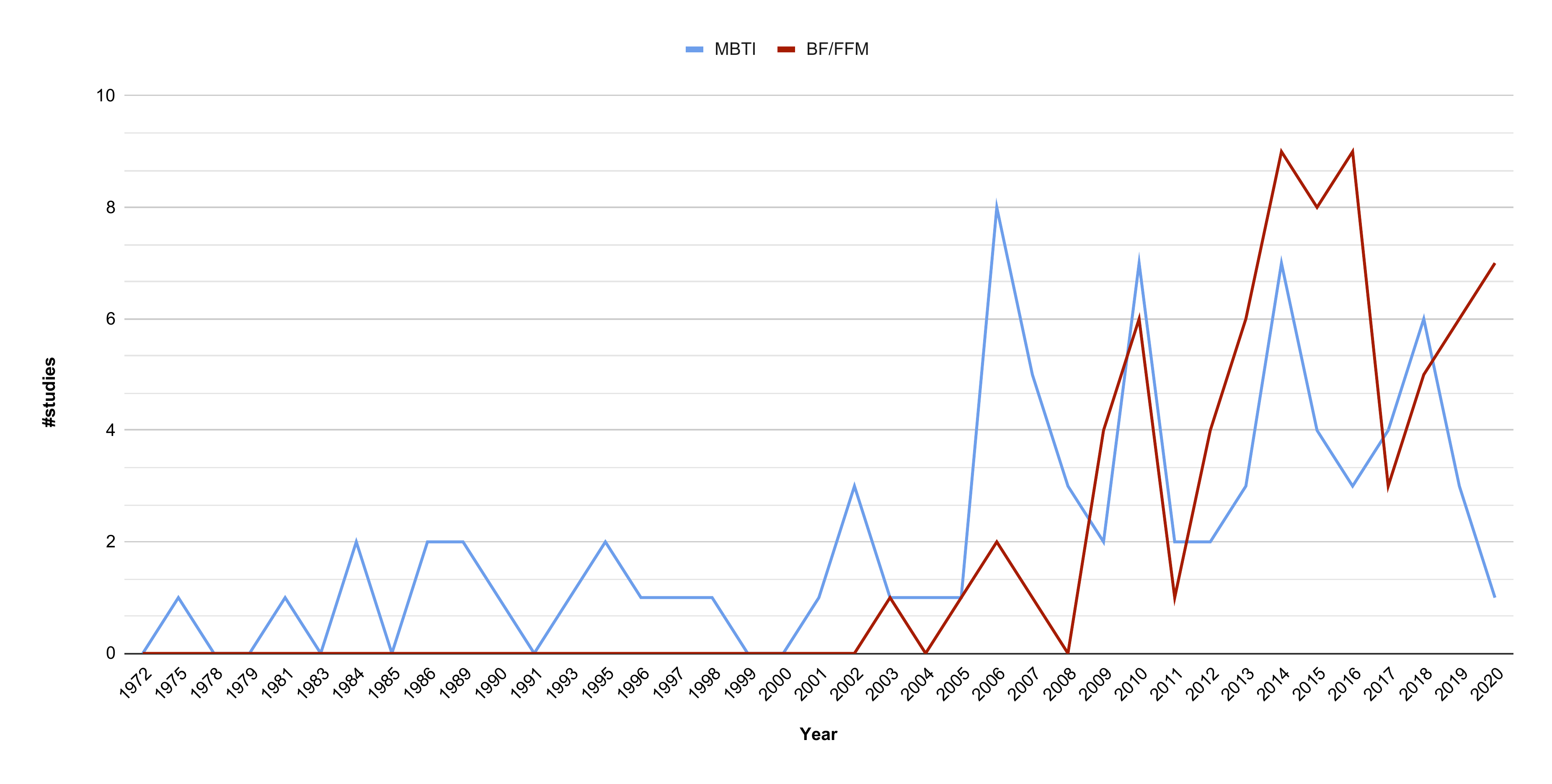}
\caption{Usage of psychometric personality instruments in SE research over the years.}
\label{fig:all-fifty-lines}
\end{figure}

Additionally, we wanted to understand if there were trends in the most adopted instruments over time. While the complete data is available in our online repository, we included only MBTI and the FFM instruments in this analysis, as none of the other instruments had comparable recent use. Figure \ref{fig:all-fifty-lines} shows how often MBTI and FFM instruments appear in SE research published from 1970 to 2020. While the first paper using MBTI was published in 1975, the first paper using FFM was published almost thirty years later, in 2003. It is possible to observe a reduction in the use of the MBTI and an increase in the use of the FFM after 2018. While we avoid risking any kind of causal inferences, the study by \cite{cruz_forty_2015} might have shed some light on criticisms regarding the use of MBTI for personality-related research \citep{boyle_myers-briggs_1995, mccrae_reinterpreting_1989}. With respect to the FFM, on the other hand, studies mainly support it as a reliable and valid theoretical background for measuring personality. For instance, the study by \cite{mccrae1987validation} provided evidence for the convergent and discriminant validity of the FFM across different measurement instruments and different observers. The study by \cite{mccrae1997personality} examined the cross-cultural generalizability of the FFM, finding support for the model in 50 cultures and languages.

Given the prevalence of the MBTI and FFM, hereafter, we discuss the use of their related instruments based on the data we extracted from the papers ranging from 2011 to 2020. Unfortunately, there was no public access to the data extracted in the mapping by \cite{cruz_forty_2015}, not allowing aggregating analyses on limitations of the usage of the instruments during this period. Furthermore, our specific focus on the instruments led us to extract more detailed information in this particular regard (\textit{cf.} Table \ref{DEF}). Nevertheless, we believe that this analysis is still meaningful, given that a more recent ten-year period actually better reflects the \textit{status quo}.

Regarding the MBTI, while there is no consensus in the literature regarding the validity of this instrument \citep{boyle_myers-briggs_1995, mccrae_reinterpreting_1989}, we found specific guidelines on how to apply it within SE research \cite{mcdonald_who_2007}. To confront the usage of the MBTI instruments with the existing guidelines, we used the extracted data on the instruments (actual name, version/bibliographic references), how they were employed (covering from administration to data interpretation), and the reported limitations. 

Not following the recommendations of these guidelines, most studies did not report any bibliographic references and explicit versions of the instrument. Furthermore, most studies also did not report anything different from ``we use x to measure personality'' regarding instruments application or ``x is widely used in SE research to measure personality'' to support the choice of the instrument. 

Following the guidelines, we extracted data from a reader's perspective looking for ``explicit details of types of test used, administration process, the qualifications of the testers'' \citep{mcdonald_who_2007} by answering the following derived questions:

\textbf{Has the study documented the participation of a qualified tester?} Only one study claims the participation of an MBTI certified practitioner to process data [S138]; however, by using surveys as empirical approach more refinements in data interpretation were limited.
    
\textbf{Are there details to justify the choice of MBTI?}
No, the choice is primarily justified by the wide use and acceptance of this instrument by previous SE studies [S151, S163, S167, S168, S169, S172, S173] or by brief claims about the validity of the instrument and professional widespread use [S155].
    
\textbf{Are there details about versions of MBTI?}
One study claims the use of \textit{Form M} or \textit{Form G}, but no references to specific versions were documented [S112, S138, S181]. Others document the use of some free tests [S129] in an unspecified version [S126]. Also, some applied the psychometric instrument through a website [S169, S172, S177] or in a printed version [S155]. Most studies document \cite{myers_mbti_1998} as a bibliographic reference when they refer to MBTI, but this reference is about the model and its theoretical foundation, not the specific psychometric instrument.
    
\textbf{How was MBTI administered?}
One study mentions having provided a participation consent form [S165]. Other studies document that instruments were self-administered by subjects through a survey empirical strategy [S130, S137, S138, S181] or in a range of time without further details (longitudinal study) [S122]. One study did not measure personality by means of MBTI but proposed a solution mapping its dimensions against adequate soft skills in requirements elicitation techniques [S179].
    
\textbf{How were the results of MBTI interpreted?}
In general, the identified studies did not report on the result interpretation. In one study, an interview followed the administration of the psychometric instrument to obtain refinements of the resulting personality traits [S112]. Additionally, one study reported the algorithm to interpret the results [S137].

In sum, we can see that despite existing literature, there is a lack of concern about how to handle personality and psychometric instruments in SE. McDonald \& Edwards' guidelines date from 2007. More recently, we had a critical review by \cite{usman_use_2019} including a sample of \cite{cruz_forty_2015} studies (see our background and related works in Subsection \ref{sec:background-personalitySE}). The results of their review are consistent with our observations, leading us to the same conclusion that there was no significant progress in improving the adoption of the MBTI in SE research over the years. The scenario is even worse when considering the lack of consensus in the literature regarding the validity of this instrument.

Regarding the FFM, despite its relevance in social sciences, we found no specific guidelines for its application by SE researchers. However, considering the coded limitations of the use of psychometric instruments reported in the studies, we observed that most studies lack details on the application of these instruments. We understand that researchers are usually restricted by document size in research papers. However, given the majority of studies mapped in journals, which are generally more extensive and detailed in terms of text, we believe that more details on how a relevant human factor as personality is handled should be provided (or at least be made available in open science repositories). 

In general, we put forward that certain basic steps should be followed when administering personality tests. It is noteworthy that the last author is a psychology researcher focused on personality-related research. Hereafter we provide some general advice adapted from the international guidelines for test use of the International Test \cite{international2001international}, which are widely accepted within the field of psychology.

\begin{itemize}
    \item The instrument used needs to show evidence of validity for the target population, and the more evidence, the better. \Ie there is a need for evidence that the test accurately assesses the personality within our population of interest. For instance, as there is limited validity evidence of personality instruments for SE professionals, one should provide arguments that allow to relate the representativeness of the sample to a population for which such evidence exists. The NEO-FFI provides evidence of the test's validity with college students and working adults in its own manual \citep{costa_mccrae_1992}. Therefore, if the study sample concerns the target population of adult software developers, a researcher could relate to such evidence arguing that adult software developers represent working adults. Researchers should look for supporting evidence of validity for the target population in the scientific literature, and new studies should be conducted when there is no such evidence.
    
    \item The instrument needs to be applied properly so that the environment does not interfere with the participant's responses. \Ie the way to apply the instrument needs to be consistent with the level of attention and comfort required to answer it. For example, consider a personality instrument with complex phrases that are difficult to understand. In that case, one must apply the test in a calm, interruption-free environment. In the context of SE professionals, this would ideally be out of the company.
    
    \item To interpret the scores obtained from the instrument, norms that are appropriate to the respondent population should be used. For example, suppose a researcher wants to interpret the Openness to Experience scores of adult woman software developers from the United States using NEO-FFI. This researcher should be aware that NEO-FFI has different norms for men and women adults from the United States, with different norms for men and women \citep{costa_mccrae_1992}. Hence, it is important to understand that norms vary depending on the test used and the population studied and that a test result is only useful when the interpretation is based on the specific norms of the target population. In this example, using the men's norms to interpret the woman's results would not be helpful.
\end{itemize}

Following these basic steps of seeking evidence of validity, standardizing the application, and using appropriate norms for the population is fundamental for a gold-standard use of psychological instruments. Hence, before using items (adjectives, descriptors, phrases, stimuli) to assess the Five Factors, there is a need for evidence that those items accurately measure the Five Factors. The instrument needs to be applied properly so that the environment does not interfere with the participant's responses, and appropriate norms need to be used to interpret the results. For example, someone will most likely introduce, or increase, measurement error by taking a German instrument, translating it into Brazilian-Portuguese, applying it in a Brazilian context, and interpreting the results based on the Germany's scores.

Furthermore, instruments may come with specific instructions, including details on their existing evidence of validity, how the instrument should be applied, and how the scores should be interpreted. In case instructions, or technical manuals, are provided, they ease the application of the international guidelines for test use by already specializing them for the instrument. For instance, the Big Five Inventory (BFI) instrument \cite{john2008paradigm} has specific advice available on the website of the Berkeley Personality Lab \footnote{https://www.ocf.berkeley.edu/~johnlab/bfi.htm}.

It is noteworthy that the advice we herein provide for the personality-related context is consistent with the more generic guidelines compiled for psychometrics in BSE \citep{graziotin2022psychometrics}.

\section{Threats to Validity}
\label{sec:results-threats}

In this section, we discuss the findings of our study regarding its threats to validity. We list the possible threats and procedures we took to mitigate those issues hereafter according to \cite{petersen_guidelines_2015}.

\textit{Theoretical validity}: with respect to our search strategy, we relied on empirically assessed guidelines to search for new evidence. Based on the set of 90 studies covering forty years of personality research in SE, we identified 6702 new studies to be analyzed in a single forward snowballing iteration using Google Scholar and included 106 additional studies. We believe that as a result, we had good coverage of the literature within the last fifty years.

Regarding study selection, the exclusion of short papers and grey literature can threaten the representativeness of the sample of selected studies. However, we adopted this strategy to prioritize complete and peer-reviewed studies. We noticed that short papers frequently did not provide the necessary information to answer our research questions during initial data extraction efforts.

Furthermore, given the huge quantity of papers to be analyzed (6702) we filtered out papers that did not include terms related to personality in the title and abstract. This decision was taken to make the study selection effort viable. We included synonyms and believe that this did not lead to relevant studies being excluded.

Finally, concerning the data extraction process, a threat can be the main control of one researcher in the mapping study execution, which can bring some bias to results in different ways. This threat was mitigated by exhausting reviewing extracted data and consensus meetings with the second author. We had support from two additional researchers in the extraction process to cover years 2017 to 2020, but the researcher that extracted data for the remaining years (2011 to 2016) reviewed their extraction. Still, the data extraction process is error-prone. To improve the reliability in this process, all extracted data is auditable and openly available to the community.

\textit{Descriptive validity}: our protocol is based on solid guidelines and an update of a comprehensive mapping study. The open coding method for answering our research questions may not help to provide an easily understandable overview. We incorporated axial coding procedures when answering research questions that primarily use open coding accordingly to Table \ref{DEF} (RQ2 and RQ3).

Regarding transparency, we documented the entire process and packaged all generated artifacts organized by the followed steps (see Figure \ref{fig:current-step}) in order to turn it available to the community. They allow further analyses and replication of our protocol. Studies that were not included are flagged with their respective exclusion criteria.

\textit{Generalizability}: The present mapping study is restricted to the dispositional personality perspective in SE research. More perspectives that could interest some target audiences may have been adopted in the SE literature. However, they were not captured by this protocol, and it is not our focus.


\section{Conclusions}
\label{sec:Conclusion}

This study aimed to provide a comprehensive overview on psychometric instruments used in SE research regarding personality. Therefore, we updated evidence of a broad existing secondary study \citep{cruz_forty_2015}. We provided a detailed protocol based on specific guidelines to assess the need for an update and to define an effective search strategy for identifying new evidence. The original secondary study included 90 studies covering 1970 to 2010, we identified 106 additional studies covering 2011 to 2020. 

By analyzing these additional studies and answering our research questions, we contribute to the Behavioral Software Engineering \citep{lenberg_behavioral_2015} body of knowledge on the following topics with respect to personality:

\begin{itemize}

    \item \textit{Status Quo}: Outlining the use of psychometric instruments in software engineering research on personality over fifty years and observing remaining discrepancies between the application of the psychometric instruments within recent SE research and existing recommendations in the literature. Within fifty years of SE research involving personality, we could not observe significant improvements. We also involved a social science researcher active with personality-related research in the analyses to provide additional advice to the SE community. 
    
    \item \textit{Common objectives}: we observed common objectives of mapped studies that use personality, employing coding procedures to provide an overview of the most studied topics. \textit{Investigate the effect of personality} in some SE activity contexts had the highest frequency, followed by \textit{characterize software engineer personality}, which aims to discover and distinguish the personality of software engineering professionals and systematizing it, mostly for mapping roles and skills.
    
    \item \textit{Limitations}: the limitations regarding the adoption of psychometric instruments are poorly reported in the mapped research. In fact, less than half of the primary studies in our set reported some limitations on the adoption of psychometric instruments.
    
    \item \textit{Theoretical constructs}: we mapped the use of psychometric instruments within recent SE research related to archetypal classes of constructs. We observed that the instruments are mainly used within the context of \textbf{actors} \textit{academic setting} and \textit{researcher} who applied the \textbf{intervention} \textit{personality traits data} to perform \textbf{activities} such as \textit{software engineer characterization}, \textit{team building}, \textit{assessment of activities execution}, \textit{pair programming}, and \textit{mapping of suitable roles}, mostly without considering a \textbf{software system} (\textit{none}) and sometimes related to \textit{class assignments} and varied \textit{tools}.
    
    \item \textit{Type of research and empirical approaches:} We provided a summary of the type of research and the empirical evaluations employing psychometric instruments. We observed that Validation research is the most common type of the mapped studies. It depicts research conducted in academic scenarios or not proposing something new and measuring in practice. With respect to the empirical evaluations, surveys and case studies are generally adopted. 
    
\end{itemize}

Overall, our study indicates that the adoption of psychometric instruments regarding personality in SE still needs to be improved. We discuss general advice from the area of social science and point readers to the recent guidelines by~\cite{graziotin2022psychometrics} on psychometrics in BSE as a first step toward improving the adoption in our discipline.

We believe that the scenario presented in our study helps to highlight this important issue and that the review and advice can help to guide the employment of psychometric instruments in SE research regarding personality.

\appendix


\section{Final list of selected primary studies}
\label{appendix:final-set}

\noindent
[S91] Omar, M., Syed-Abdullah, S.-L., et al. (2011). \emph{Developing a Team Performance Prediction Model: A Rough Sets Approach}. Informatics Engineering and Information Science (pp. 691–705). Springer.
\bigskip

\noindent
[S92] Salleh, N., Mendes, E., et al. (2011). \emph{The effects of openness to experience on pair programming in a higher education context}. 2011 24th IEEE-CS Conference on Software Engineering Education and Training (CSEET), 149–158.
\bigskip

\noindent
[S93] Raza, A., UlMustafa, Z., et al. (2011). \emph{Personality Dimensions and Temperaments of Engineering Professors and Students – A Survey}. Journal of Computing, 3(12), 13–20.
\bigskip

\noindent
[S94] Yilmaz, M.,\& OConnor, R. V. (2012). \emph{Towards the Understanding and Classification of the Personality Traits of Software Development Practitioners: Situational Context Cards Approach}. 2012 38th Euromicro Conference on Software Engineering and Advanced Applications, 400–405.
\bigskip

\noindent
[S95] Gómez, M. N., Acuña, S. T., et al. (2012). \emph{How Does the Extraversion of Software Development Teams Influence Team Satisfaction and Software Quality?: A Controlled Experiment}. International Journal of Human Capital and Information Technology Professionals (IJHCITP), 3(4), 11–24.
\bigskip

\noindent
[S96] Sfetsos, P., Adamidis, P., et al. (2012). \emph{Investigating the Impact of Personality and Temperament Traits on Pair Programming: A Controlled Experiment Replication}. 2012 Eighth International Conference on the Quality of Information and Communications Technology, 57–65.
\bigskip

\noindent
[S97] Stylianou, C.,\& Andreou, A. S. (2012). \emph{A Multi-objective Genetic Algorithm for Software Development Team Staffing Based on Personality Types}. Artificial Intelligence Applications and Innovations (pp. 37–47). Springer.
\bigskip

\noindent
[S98] Stylianou, C., Gerasimou, S., et al. (2012). \emph{A Novel Prototype Tool for Intelligent Software Project Scheduling and Staffing Enhanced with Personality Factors}. 2012 IEEE 24th International Conference on Tools with Artificial Intelligence, 1, 277–284. https://doi.org/10.1109/ICTAI.2012.45
\bigskip

\noindent
[S99] Montequín, V. R., Balsera, J. V., et al. (2012). \emph{Using Myers-Briggs Type Indicator (MBTI) as a Tool for Setting up Student Teams for Information Technology Projects}. Journal of Information Technology and Application in Education, 1, 7.
\bigskip

\noindent
[S100] Mair, C., Martincova, M., et al. (2012). \emph{An Empirical Study of Software Project Managers Using a Case-Based Reasoner}. 2012 45th Hawaii International Conference on System Sciences, 1030–1039.
\bigskip

\noindent
[S101] Gorla, N., Chiravuri, A., et al. (2013). \emph{Effect of personality type on structured tool comprehension performance}. Requirements Engineering, 18(3), 281–292.
\bigskip

\noindent
[S102] Koroutchev, K., Acuña, S. T., et al. (2013). \emph{The Social Environment as a Determinant for the Impact of the Big Five Personality Factors and the Group’s Performance}. International Journal of Human Capital and Information Technology Professionals (IJHCITP), 4(1), 1–8.
\bigskip

\noindent
[S103] Sfetsos, P, Adamidis, P., et al. (2013). \emph{Heterogeneous Personalities Perform Better in Pair Programming: The Results of a Replication Study}. Software Quality Professional Magazine, 15(4)
\bigskip

\noindent
[S104] Bishop, D. (2013). \emph{Personality Theory as a Predictor for Agile Preference}. MWAIS 2013 Proceedings.
\bigskip

\noindent
[S105] Kanij, T., Merkel, R., et al. (2013). \emph{An empirical study of the effects of personality on software testing}. 2013 26th International Conference on Software Engineering Education and Training (CSEE T), 239–248.
\bigskip

\noindent
[S106] Radhakrishnan, P.,\& Kanmani, S. (2013). \emph{Improvement of programming skills using pair programming by boosting extraversion and openness to experience}. International Journal of Teaching and Case Studies, 4(1), 13–35.
\bigskip

\noindent
[S107] Varathan, K. D.,\& Thiam, L. T. (2013) \emph{Mining Facebook in Identifying Software Engineering Students’ Personality and Job Matching}. ISSN:2188-272X – The Asian Conference on Society, Education and Technology 2013 – Official Conference Proceedings.
\bigskip

\noindent
[S108] Karapıçak, Ç. M.,\& Demirörs, O. (2013). \emph{A Case Study on the Need to Consider Personality Types for Software Team Formation}. Software Process Improvement and Capability Determination (pp. 120–129). Springer. 
\bigskip

\noindent
[S109] Thiti Phiriyayotha, Siriluck Rotchanakitumnuai. (2013). \emph{Data Warehouse Implementation Success Factors and the Impact of Leadership and Personality on the Relationship between Success Factors}. Journal of Business and Economics, 4(10), 948–956.
\bigskip

\noindent
[S110] Luse, A., McElroy, J. C., et al. (2013). \emph{Personality and cognitive style as predictors of preference for working in virtual teams}. Computers in Human Behavior, 29(4), 1825–1832.
\bigskip

\noindent
[S111] Martínez, L. G., Licea, G., et al. (2013). \emph{Using MatLab’s fuzzy logic toolbox to create an application for RAMSET in software engineering courses}. Computer Applications in Engineering Education, 21(4), 596–605.
\bigskip

\noindent
[S112] Raza, A., Capretz, L.F., Ul-Mustafa, Z., 2014. \emph{Personality Profiles of Software Engineers and Their Software Quality Preferences}. International Journal of Information Systems and Social Change (IJISSC) 5, 77–86.
\bigskip

\noindent
[S113] Salleh, N., Mendes, E., Grundy, J., 2014. \emph{Investigating the effects of personality traits on pair programming in a higher education setting through a family of experiments}. Empirical Software Engineering 19, 714–752.
\bigskip

\noindent
[S114] Bishop, D., Deokar, A., 2014. \emph{Toward an Understanding of Preference for Agile Software Development Methods from a Personality Theory Perspective}. 47th Hawaii International Conference on System Sciences, pp. 4749–4758.
\bigskip

\noindent
[S115] Venkatesan, V., Sankar, A., 2014. \emph{Investigation of Students Personality on Pair Programming to Enhance the Learning Activity in the Academia}. Journal of Computer Science 10, 2020–2028.
\bigskip

\noindent
[S116] Licorish, S.A., MacDonell, S.G., 2014. \emph{Personality profiles of global software developers}, in: Proceedings of the 18th International Conference on Evaluation and Assessment in Software Engineering, EASE ’14. Association for Computing Machinery, New York, NY, USA, pp. 1–10.
\bigskip

\noindent
[S117] Gómez, M.N., Acuña, S.T., 2014. \emph{A replicated quasi-experimental study on the influence of personality and team climate in software development}. Empir Software Eng 19, 343–377.
\bigskip

\noindent
[S118] Kosti, M.V., Feldt, R., Angelis, L., 2014. \emph{Personality, emotional intelligence and work preferences in software engineering: An empirical study}. Information and Software Technology 56, 973–990.
\bigskip

\noindent
[S119] Okike, E.U., 2014. \emph{Bipolar Factor and Systems Analysis skills of Student Computing Professionals at University of Botswana, Gaborone}. International Journal of Advanced Computer Science and Applications (IJACSA) 5.
\bigskip

\noindent
[S120] Okike, E.U., 2014. \emph{Investigating Students’ Achievements in Computing Science Using Human Metric.} International Journal of Advanced Computer Science and Applications (IJACSA) 5.
\bigskip

\noindent
[S121] Yilmaz, M., O’Connor, R.V., Clarke, P., 2014. \emph{An Exploration of Individual Personality Types in Software Development}, in: Barafort, B., O’Connor, R.V., Poth, A., Messnarz, R. (Eds.), Systems, Software and Services Process Improvement, Communications in Computer and Information Science. Springer, Berlin, Heidelberg, pp. 111–122.
\bigskip

\noindent
[S122] Kruck, S.E., Sendall, P., Ceccucci, W., Peslak, A., Hunsinger, S., 2014. \emph{Does Personality Play a Role in Computer Information Systems Performance?}. Issues in Information Systems 15, 383–392.
\bigskip

\noindent
[S123] Ezekiel U. Okike and Olanrewaju A. Amoo, 2014. \emph{Problem Solving and Decision Making: Consideration of Individual Differences in Computer programming Skills Using Myers Briggs Type Indicator (MBTI) and Chidamber and Kemerer Java Metrics (CKJM)}. 7, 8.
\bigskip

\noindent
[S124] Papatheocharous, E., Belk, M., Nyfjord, J., Germanakos, P., Samaras, G., 2014. \emph{Personalised continuous software engineering}, in: Proceedings of the 1st International Workshop on Rapid Continuous Software Engineering, RCoSE 2014. Association for Computing Machinery, New York, NY, USA, pp. 57–62.
\bigskip

\noindent
[S125] Gramß, D., Frank, T., Rehberger, S., Vogel-Heuser, B., 2014. \emph{Female characteristics and requirements in software engineering in mechanical engineering}, in: 2014 International Conference on Interactive Collaborative Learning (ICL). Presented at the 2014 International Conference on Interactive Collaborative Learning (ICL), pp. 272–279.
\bigskip

\noindent
[S126] Okike, E.U., 2014. \emph{A Code Level Based Programmer Assessment and Selection Criterion Using Metric Tools}. International Journal of Advanced Computer Science and Applications (IJACSA) 5.
\bigskip

\noindent
[S127] Vitó Ferreira, N.N., Langerman, J.J., 2014. \emph{The correlation between personality type and individual performance on an ICT Project}, in: 2014 9th International Conference on Computer Science Education. Presented at the 2014 9th International Conference on Computer Science Education, pp. 425–430.
\bigskip

\noindent
[S128] Huang, F., Liu, B., Song, Y., Keyal, S., 2014. \emph{The links between human error diversity and software diversity: Implications for fault diversity seeking}. Science of Computer Programming 89, 350–373.
\bigskip

\noindent
[S129] Ribaud, V., \& Saliou, P. (2015). \emph{Relating ICT Competencies with Personality Types}. Systems, Software and Services Process Improvement (pp. 295–302). Springer International Publishing.
\bigskip

\noindent
[S130] Branco, D. T. M. C., Oliveira, E. C. C. de, Galvão, L., Prikladnicki, R., \& Conte, T. (2015). \emph{An Empirical Study about the Influence of Project Manager Personality in Software Project Effort}. 102–113.
\bigskip

\noindent
[S131] Yilmaz, M., \& O’Connor, R. V. (2015). \emph{Understanding personality differences in software organisations using Keirsey temperament sorter}. IET Software, 9(5), 129–134.
\bigskip

\noindent
[S132] Kanij, T., Merkel, R., \& Grundy, J. (2015). \emph{An Empirical Investigation of Personality Traits of Software Testers}. 2015 IEEE/ACM 8th International Workshop on Cooperative and Human Aspects of Software Engineering, 1–7.
\bigskip

\noindent
[S133] Acuña, S. T., Gómez, M. N., Hannay, J. E., Juristo, N., \& Pfahl, D. (2015). \emph{Are team personality and climate related to satisfaction and software quality? Aggregating results from a twice replicated experiment}. Information and Software Technology, 57, 141–156.
\bigskip

\noindent
[S134] Licorish, S. A., \& MacDonell, S. G. (2015). \emph{Communication and personality profiles of global software developers}. Information and Software Technology, 64, 113–131.
\bigskip

\noindent
[S135] Karimi, Z., Baraani-Dastjerdi, A., Ghasem-Aghaee, N., \& Wagner, S. (2015). \emph{Influence of Personality on Programming Styles an Empirical Study}. Journal of Information Technology Research (JITR), 8(4), 38–56.
\bigskip

\noindent
[S136] Monaghan, C., Bizumic, B., Reynolds, K., Smithson, M., Johns-Boast, L., \& van Rooy, D. (2015). \emph{Performance of student software development teams: The influence of personality and identifying as team members}. European Journal of Engineering Education, 40(1), 52–67.
\bigskip

\noindent
[S137] Dargis, M., Finke, A., \& Penicina, L. (2015). \emph{Relationship between Personality Types Conceptualized by C. G. Jung and Latvian IT Specialist Preferences}. Complex Systems Informatics and Modeling Quarterly, 4, 1–11.
\bigskip

\noindent
[S138] Capretz, L. F., Varona, D., \& Raza, A. (2015). \emph{Influence of personality types in software tasks choices}. Computers in Human Behavior, 52, 373–378.
\bigskip

\noindent
[S139] Soomro, A. B., Salleh, N., \& Nordin, A. (2015). \emph{How personality traits are interrelated with team climate and team performance in software engineering? A preliminary study}. 2015 9th Malaysian Software Engineering Conference (MySEC), 259–265.
\bigskip

\noindent
[S140] Suslow, W., Kowalczyk, J., Statkiewicz, M., Boinska, M., \& Nowak, J. (2015). \emph{Psychosocial Correlates of Software Designers’ Professional Aptitude}. International Journal of Advanced Computer Science and Applications (IJACSA), 6(8), Article 8.
\bigskip

\noindent
[S141] McDermott, R., Daniels, M., \& Cajander, Å. (2015). \emph{Perseverance Measures and Attainment in First Year Computing Science Students}. Proceedings of the 2015 ACM Conference on Innovation and Technology in Computer Science Education, 302–307.
\bigskip

\noindent
[S142] Murukannaiah, P. K., Ajmeri, N., et al. (2016). \emph{Acquiring Creative Requirements from the Crowd: Understanding the Influences of Personality and Creative Potential in Crowd RE}. 2016 IEEE 24th International Requirements Engineering Conference (RE), 176–185.
\bigskip

\noindent
[S143] Ostberg, J.-P., Wagner, S., et al. (2016). \emph{Does Personality Influence the Usage of Static Analysis Tools? An Explorative Experiment}. 2016 IEEE/ACM Cooperative and Human Aspects of Software Engineering (CHASE), 75–81.
\bigskip

\noindent
[S144] Mujkanovic, A., \& Bollin, A. (2016). \emph{Improving Learning Outcomes through Systematic Group Reformation—The Role of Skills and Personality in Software Engineering Education}. 2016 IEEE/ACM Cooperative and Human Aspects of Software Engineering (CHASE), 97–103.
\bigskip

\noindent
[S145] Eckhardt, A., Laumer, S., et al. (2016). \emph{The Effect of Personality on IT Personnel’s Job-Related Attitudes: Establishing a Dispositional Model of Turnover Intention across IT Job Types}. Journal of Information Technology, 31(1), 48–66.
\bigskip

\noindent
[S146] Karimi, Z., Baraani-Dastjerdi, A., et al. (2016). \emph{Using Personality Traits to Understand the Influence of Personality on Computer Programming: An Empirical Study}. Journal of Cases on Information Technology (JCIT), 18(1), 28–48.
\bigskip

\noindent
[S147] Hsu, W.-C., \& Barnes, G. M. (2016). \emph{Are computer information technology majors different than computer science majors in personality, learning style, or academic performance?}. Journal of Computing Sciences in Colleges, 31(4), 100–107.
\bigskip

\noindent
[S148] Agung, A. A. G., \& Yuniar, I. (2016). \emph{Personality assessment website using DISC: A case study in information technology school}. 2016 International Conference on Information Management and Technology (ICIMTech), 72–77.
\bigskip

\noindent
[S149] Rastogi, A., \& Nagappan, N. (2016). \emph{On the Personality Traits of GitHub Contributors}. 2016 IEEE 27th International Symposium on Software Reliability Engineering (ISSRE), 77–86.
\bigskip

\noindent
[S150] Karimi, Z., Baraani-Dastjerdi, A., et al. (2016). \emph{Links between the personalities, styles and performance in computer programming}. Journal of Systems and Software, 111, 228–241.
\bigskip

\noindent
[S151] Gilal, A. R., Jaafar, J., et al. (2016). \emph{Balancing the Personality of Programmer: Software Development Team Composition}. Malaysian Journal of Computer Science, 29(2), 145–155.
\bigskip

\noindent
[S152] Paruma Pabón, O. H., González, F. A., et al. (2016). \emph{Finding Relationships between Socio-Technical Aspects and Personality Traits by Mining Developer E-mails}. 2016 IEEE/ACM Cooperative and Human Aspects of Software Engineering (CHASE), 8–14.
\bigskip

\noindent
[S153] Aeron, S., \& Pathak, S. (2016). \emph{Personality composition in Indian software teams and its relationship to social cohesion and task cohesion}. International Journal of Indian Culture and Business Management, 13(3), 267–287.
\bigskip

\noindent
[S154] Yilmaz, M., Yilmaz, M., et al. (2016). \emph{A Gamification Approach to Improve the Software Development Process by Exploring the Personality of Software Practitioners}. Software Process Improvement and Capability Determination (pp. 71–83). Springer International Publishing. \bigskip

\noindent
[S155] John H. Reynolds, D. Robert Adams, et al. (2016). \emph{The Personality of a Computing Major: It Makes a Difference}. EDSIGCON Proceeding 2016, 2. 
\bigskip

\noindent
[S156] Kosti, M. V., Feldt, R., et al. (2016). \emph{Archetypal personalities of software engineers and their work preferences: A new perspective for empirical studies}. Empirical Software Engineering, 21(4), 1509–1532.
\bigskip

\noindent
[S157] Xia, X., Lo, D., et al. (2017). \emph{Personality and Project Success: Insights from a Large-Scale Study with Professionals}. 2017 IEEE International Conference on Software Maintenance and Evolution (ICSME), 318–328.
\bigskip

\noindent
[S158] Gilal, A. R., Jaafar, J., et al. (2017). \emph{Making Programmer Effective for Software Development Teams: An Extended Study}. Journal of Information Science and Engineering, 33(6), 17.
\bigskip

\noindent
[S159] Rehman, M., Safdar, S., et al. (2017). \emph{Personality traits and knowledge sharing behavior of software engineers}. 2017 8th International Conference on Information Technology (ICIT), 6–11. 
\bigskip

\noindent
[S160] Yilmaz, M., O’Connor, R. V., et al. (2017). \emph{An examination of personality traits and how they impact on software development teams}. Information and Software Technology, 86, 101–122.
\bigskip

\noindent
[S161] Çelikten, A.,\& Çetin, A. (2017). \emph{Assigning Product Development Roles to Software Engineers Based on Personality Types and Skills}. Academic Journal of Science, 7(3), 475–486.
\bigskip

\noindent
[S162] Tunio, M. Z., Luo, H., et al. (2017). \emph{Impact of Personality on Task Selection in Crowdsourcing Software Development: A Sorting Approach}. IEEE Access, 5, 18287–18294.
\bigskip

\noindent
[S163] Gilal, A. R., Jaafar, J., et al. (2017). \emph{Suitable Personality Traits for Learning Programming Subjects: A Rough-Fuzzy Model}. International Journal of Advanced Computer Science and Applications (IJACSA), 8(8), Article 8.
\bigskip

\noindent
[S164] Barroso, A. S., Madureira, J. S., et al. (2017). \emph{Relationship between Personality Traits and Software Quality—Big Five Model vs. Object-oriented Software Metrics}. 63–74.
\bigskip

\noindent
[S165] Gilal, A. R., Omar, M., et al. (2017). \emph{Software development team composition: Personality types of programmer and complex network}. 153–159.
\bigskip

\noindent
[S166] Jevsikova, T., Dagienė, V., et al. (2018). \emph{On Preferences of Novice Software Engineering Students: Temperament Style and Attitudes Towards Programming Activities}. Fundamentals of Computer Science and Software Engineering (pp. 101–113). Springer International Publishing.
\bigskip

\noindent
[S167] Tunio, M. Z., Luo, H., Wang, C.,\& Zhao, F. (2018). \emph{Crowdsourcing Software Development: Task Assignment Using PDDL Artificial Intelligence Planning}. Journal of Information Processing Systems, 14(1), 129–139.
\bigskip

\noindent
[S168] Tunio, M. Z., Luo, H., Wang, C., Zhao, F., et al. (2018). \emph{Task Assignment Model for Crowdsourcing Software Development: TAM}. Journal of Information Processing Systems, 14(3), 621–630.
\bigskip

\noindent
[S169] Perez-Gonzalez, H. G., Nunez-Varela, A., et al. (2018). \emph{Investigating the Effects of Personality on Software Design in a Higher Education Setting Through an Experiment}. 2018 6th International Conference in Software Engineering Research and Innovation (CONISOFT), 72–78. \bigskip

\noindent
[S170] Sánchez, M. G. T., Cachero, C., et al. (2018). \emph{Evaluating the Effect of Developers’ Personality and Productivity on their Intention to Use Model-Driven Web Engineering Techniques: An Exploratory Observational Study}. Journal of Web Engineering, 483–526.
\bigskip

\noindent
[S171] Anderson, G., Keith, M. J., et al. (2018). \emph{The Effect of Software Team Personality Composition on Learning and Performance: Making the “Dream” Team}. https://doi.org/10.24251/HICSS.2018.059
\bigskip

\noindent
[S172] Garousi, V.,\& Tarhan, A. (2018). \emph{Investigating the Impact of Team Formation by Introversion/Extraversion in Software Projects}. Balkan Journal of Electrical and Computer Engineering, 6(2), 132–140.
\bigskip

\noindent
[S173] Gilal, A. R., Jaafar, J., et al. (2018). \emph{Finding an effective classification technique to develop a software team composition model}. Journal of Software: Evolution and Process, 30(1), e1920.
\bigskip

\noindent
[S174] Shuto, M., Washizaki, H., et al. (2018). \emph{Personality and Learning Effectiveness of Teams in Information Systems Education Courses}. EAI Endorsed Transactions on E-Learning, 5(17).
\bigskip

\noindent
[S175] Ting, T. L.,\& Varathan, K. D. (2018).\emph{JOB RECOMMENDATION USING FACEBOOK PERSONALITY SCORES}. Malaysian Journal of Computer Science, 31(4), 311–331.
\bigskip

\noindent
[S176] Chowdhury, S., Walter, C., et al. (2018). \emph{Toward Increasing Collaboration Awareness in Software Engineering Teams}. 2018 IEEE Frontiers in Education Conference (FIE), 1–9.
\bigskip

\noindent
[S177] Barroso, A. S., de J. Prado, K. H., et al. (2019). \emph{How personality traits influences quality of software developed by students}. Proceedings of the XV Brazilian Symposium on Information Systems, 1–8.
\bigskip

\noindent
[S178] Mellblom, E., Arason, I., et al. (2019). \emph{The Connection Between Burnout and Personality Types in Software Developers}. IEEE Software, 36(5), 57–64.
\bigskip

\noindent
[S179] Khan, A., Gilal, A. R., et al. (2019). \emph{Does Software Requirement Elicitation and Personality make any relation?}. Journal of Advanced Research in Dynamic and Control Systems, Volume 11(08-Special Issue), 1162–1168.
\bigskip

\noindent
[S180] Calefato, F., Lanubile, F., et al. (2019). \emph{A large-scale, in-depth analysis of developers’ personalities in the Apache ecosystem}. Information and Software Technology, 114, 1–20.
\bigskip

\noindent
[S181] Hasan, A., Moin, S., et al. (2019). \emph{Prediction of Personality Profiles in the Pakistan Software Industry–A Study}. Psych, 1(1), 320–330.
\bigskip

\noindent
[S182] Qamar, N.,\& Malik, A. A. (2019). \emph{Birds of a Feather Gel Together: Impact of Team Homogeneity on Software Quality and Team Productivity}. IEEE Access, 7, 96827–96840.
\bigskip

\noindent
[S183] Wyrich, M., Graziotin, D., et al. (2019). \emph{A theory on individual characteristics of successful coding challenge solvers}. PeerJ Computer Science, 5, e173.
\bigskip

\noindent
[S184] Akarsu, Z., Orgun, P., et al. (2019). \emph{Assessing Personality Traits in a Large Scale Software Development Company: Exploratory Industrial Case Study}. Systems, Software and Services Process Improvement, 192–206.
\bigskip

\noindent
[S185] Celikten, A., Kurt, E., et al. (2019). \emph{A Decision Support System for Role Assignment in Software Project Management with Evaluation of Personality Types}. Artificial Intelligence and Applied Mathematics in Engineering Problems (pp. 200–210). Springer International Publishing.
\bigskip

\noindent
[S186] Khan, I. A., Khan, A., et al. (2019). \emph{Urdu Translation: The Validation and Reliability of the 120-Item Big Five IPIP Personality Scale}. Current Psychology, 38(6), 1530–1541.
\bigskip

\noindent
[S187] Gomes, A., Silva, M., et al. (2020). \emph{Evaluating the Relationship of Personality and Teamwork Quality in the Context of Agile Software Development}. SEEK 2020 Proceedings, 6.
\bigskip

\noindent
[S188] Chiu, Y.-T., Day, M.-Y., et al. (2020). \emph{Exploring impacts of project leaders’ written expressions in virtual and fluid projects: The role of personality and emotion}. International Research Workshop on IT Project Management 2020. https://aisel.aisnet.org/irwitpm2020/5
\bigskip

\noindent
[S189] Amin, A., Basri, S., et al. (2020). \emph{The impact of personality traits and knowledge collection behavior on programmer creativity}. Information and Software Technology, 128, 106405.
\bigskip

\noindent
[S190] Akarsu, Z.,\& Yilmaz, M. (2020). \emph{Managing the social aspects of software development ecosystems: An industrial case study on personality}. Journal of Software: Evolution and Process, 32(11), e2277.
\bigskip

\noindent
[S191] Russo, D.,\& Stol, K.-J. (2020). \emph{Gender Differences in Personality Traits of Software Engineers}. IEEE Transactions on Software Engineering, 48(3), 819–834.
\bigskip

\noindent
[S192] Doungsa-ard, C.,\& Chaiwon, V. (2020). \emph{The Software Engineering Position Mapping From Personality Traits}. 2020 Joint International Conference on Digital Arts, Media and Technology with ECTI Northern Section Conference on Electrical, Electronics, Computer and Telecommunications Engineering (ECTI DAMT NCON), 194–199.
\bigskip

\noindent
[S193] Demir, Ö.,\& Seferoglu, S. S. (2020). \emph{The Effect of Determining Pair Programming Groups According to Various Individual Difference Variables on Group Compatibility, Flow, and Coding Performance}. Journal of Educational Computing Research, 59(1), 41–70.
\bigskip

\noindent
[S194] Papoutsoglou, M., Kapitsaki, G. M., et al. (2020). \emph{Modeling the effect of the badges gamification mechanism on personality traits of Stack Overflow users}. Simulation Modelling Practice and Theory, 105, 102157.
\bigskip

\noindent
[S195] Vishnubhotla, S. D., Mendes, E., et al. (2020). \emph{Investigating the relationship between personalities and agile team climate of software professionals in a telecom company}. Information and Software Technology, 126, 106335.
\bigskip

\noindent
[S196] Tehlan, K., Chakraverty, S., et al. (2020). \emph{A genetic algorithm-based approach for making pairs and assigning exercises in a programming course}. Computer Applications in Engineering Education, 28(6), 1708–1721.

\pagebreak


\section{Dictionary of coded constructs related to SE}
\label{appendix:coded-constructs}

\begin{table}[h]
\centering
\caption{Coded constructs in class Actor}
\label{tab:app-actor}
\resizebox{\textwidth}{!}{%
\begin{tabular}{|l|l|l|r|}
\hline
\rowcolor[HTML]{EFEFEF} 
\multicolumn{1}{|c|}{\cellcolor[HTML]{EFEFEF}\textbf{Actor coded construct}} & \multicolumn{1}{c|}{\cellcolor[HTML]{EFEFEF}\textbf{Description}} & \multicolumn{1}{c|}{\cellcolor[HTML]{EFEFEF}\textbf{Studies}} & \multicolumn{1}{c|}{\cellcolor[HTML]{EFEFEF}\textbf{Count}} \\ \hline
academic setting & \begin{tabular}[c]{@{}l@{}}The study has an educational purpose \\ or is applied in an academic setting due\\ to scope limitations\end{tabular} & \begin{tabular}[c]{@{}l@{}}S91, S92, S95, S96, S99, S101, \\ S102, S103, S105, S106, S107, S110,\\ S111, S113, S115, S120, S122, S125, \\ S126, S133, S135, S136, S140, S141, \\ S143, S144, S146, S147, S148, S150, \\ S151, S155, S158, S162, S163, S165, \\ S166, S169, S171, S172, S173, S174, \\ S176, S177, S182, S193, S196\end{tabular} & 47 \\ \hline
researcher & \begin{tabular}[c]{@{}l@{}}The study author(s) act primarily for \\ investigative purposes\end{tabular} & \begin{tabular}[c]{@{}l@{}}S93, S104, S108, S112, S114, S116, \\ S117, S118, S119, S121, S123, S124, \\ S128, S129, S131, S132, S134, S137, \\ S138, S139, S145, S149, S152, S156, \\ S159, S161, S167, S168, S170, S175, \\ S178, S179, S180, S181, S183, S185, \\ S186, S187, S188, S189, S191, S192, \\ S194\end{tabular} & 43 \\ \hline
industrial setting & \begin{tabular}[c]{@{}l@{}}The study is applied in a industrial \\ setting\end{tabular} & \begin{tabular}[c]{@{}l@{}}S109, S130, S142, S153, S157, S160, \\ S164, S184, S190, S195\end{tabular} & 10 \\ \hline
organization & \begin{tabular}[c]{@{}l@{}}The scope of the study is not clear \\ (where does the study data come \\ from?). The code was adopted to \\ be comprehensive.\end{tabular} & S97, S98, S100, S127 & 4 \\ \hline
practitioner & \begin{tabular}[c]{@{}l@{}}The study is clearly applied and focu-\\ sed on practioners\end{tabular} & S154 & 1 \\ \hline
software development team & \begin{tabular}[c]{@{}l@{}}A software developent team was the\\  interventor in the study\end{tabular} & S94 & 1 \\ \hline
\end{tabular}%
}
\end{table}

\begin{table}[]
\centering
\caption{Coded constructs in class Intervention}
\label{tab:app-intervention}
\resizebox{\textwidth}{!}{%
\begin{tabular}{|l|l|l|r|}
\hline
\rowcolor[HTML]{EFEFEF} 
\multicolumn{1}{|c|}{\cellcolor[HTML]{EFEFEF}\textbf{Intervention coded construct}} & \multicolumn{1}{c|}{\cellcolor[HTML]{EFEFEF}\textbf{Description}} & \multicolumn{1}{c|}{\cellcolor[HTML]{EFEFEF}\textbf{Studies}} & \multicolumn{1}{c|}{\cellcolor[HTML]{EFEFEF}\textbf{Count}} \\ \hline
personality traits data & \begin{tabular}[c]{@{}l@{}}The study has data on personality traits \\ measured by some psychometric instrument\end{tabular} & \begin{tabular}[c]{@{}l@{}}S91, S92, S93, S94, S95, S96, \\S97, S98, S99, S100, S101, S102,\\ S103, S104, S105, S107, S108, S109,  \\S110, S111, S112, S113, S114, S115, \\S116, S117, S118, S119, S120, S122, \\S123, S124, S125, S126, S127, S128,\\ S129, S130, S131, S132, S133, S134, \\S135, S136, S137, S138, S139, S140, \\S141, S142, S143, S144, S145, S146,\\S147, S148, S149, S150, S151, S152, \\S153, S155, S156, S157, S158, S159, \\S160, S161, S162, S163, S164, S165, \\S166, S167, S168, S169, S170, S171, \\S172, S173, S174,  S176, S177, S178, \\S179, S180, S181, S182, S183, S184,\\  S185, S186, S187, S188, S189, S190, \\S191, S192, S193, S194, S195, S196\end{tabular} & 102 \\ \hline
pair programming & \begin{tabular}[c]{@{}l@{}}The study adopted pair programming to \\ assess some impact\end{tabular} & S106 & 1 \\ \hline
psychometric instrument & \begin{tabular}[c]{@{}l@{}}The study adopted a psychometric instrument\\  to be adapted/validated\end{tabular} & S121 & 1 \\ \hline
social media personality traits data & \begin{tabular}[c]{@{}l@{}}The study has data on personality measure \\ using social media data as source\end{tabular} & S175 & 1 \\ \hline
virtual environment to measure personality & \begin{tabular}[c]{@{}l@{}}The study used an virtual enviroment to mea-\\ sure personality as intervention\end{tabular} & S154 & 1 \\ \hline
software development team & \begin{tabular}[c]{@{}l@{}}A software developent team was the interv-\\ entor in the study\end{tabular} & S94 & 1 \\ \hline
\end{tabular}%
}
\end{table}

\begin{table}[]
\centering
\caption{Coded constructs in class Activity}
\label{tab:app-activity}
\resizebox{\textwidth}{!}{%
\begin{tabular}{|l|l|l|r|}
\hline
\rowcolor[HTML]{EFEFEF} 
\multicolumn{1}{|c|}{\cellcolor[HTML]{EFEFEF}\textbf{Activity coded construct}} & \multicolumn{1}{c|}{\cellcolor[HTML]{EFEFEF}\textbf{Description}} & \multicolumn{1}{c|}{\cellcolor[HTML]{EFEFEF}\textbf{Studies}} & \multicolumn{1}{c|}{\cellcolor[HTML]{EFEFEF}\textbf{Count}} \\ \hline
characterization of software engineer & \begin{tabular}[c]{@{}l@{}}The study characterizes, in some extent,\\ the software engineering professional. \\ It means: comparison of SE and other \\ professionals, discover (and/or compare)\\ personalities for some purpose.\end{tabular} & \begin{tabular}[c]{@{}l@{}}S93, S108, S112, S116, S118, S119,\\ S127, S129, S131, S132, S134, S137,\\ S138, S147, S148, S149, S152, S155,\\ S156, S160, S165, S179, S180, S184,\\ S191, S194, S195\end{tabular} & 27 \\ \hline
team building & \begin{tabular}[c]{@{}l@{}}The main activity of the study is to build\\ software development teams with more\\ than two members\end{tabular} & \begin{tabular}[c]{@{}l@{}}S91, S94, S95, S97, S98, S99,\\ S102, S111, S124, S133, S144, S153,\\  S158, S171, S172, S174\end{tabular} & 16 \\ \hline
assessment of activities execution & \begin{tabular}[c]{@{}l@{}}The main activity of the study is to assess\\ software artifacts and processes in a soft-\\ ware activity\end{tabular} & \begin{tabular}[c]{@{}l@{}}S101, S117, S120, S122, S125, S126,\\ S143, S159, S169, S182\end{tabular} & 10 \\ \hline
pair programming & \begin{tabular}[c]{@{}l@{}}The main activity of the study is to build\\ pair programming teams\end{tabular} & S92, S96, S103, S113, S115, S193, S196 & 7 \\ \hline
mapping of suitable roles & \begin{tabular}[c]{@{}l@{}}The main activity of the study is to map-\\ ping software engineer roles (developer,\\ tester, project manager, etc.)\end{tabular} & S107, S161, S175, S185, S192 & 5 \\ \hline
investigate team performance & \begin{tabular}[c]{@{}l@{}}The main activity of study is inves-\\ tigate performance of a software team\end{tabular} & S136, S139, S173, S176, S190 & 5 \\ \hline
inquiry on programmer performance & \begin{tabular}[c]{@{}l@{}}The main activity is assess program-\\ mer performance in coding activities\end{tabular} & S123, S128, S146, S163 & 4 \\ \hline
project management activities & \begin{tabular}[c]{@{}l@{}}The main activity is related to soft-\\ ware project management activities\end{tabular} & S100, S130, S188 & 3 \\ \hline
task selection & \begin{tabular}[c]{@{}l@{}}The main activity is related to selection\\ of tasks to development\end{tabular} & S162, S167, S168 & 3 \\ \hline
\end{tabular}%
}
\end{table}

\begin{table}[]
\centering
\caption{Coded constructs in class Software System}
\label{tab:app-ss}
\resizebox{\textwidth}{!}{%
\begin{tabular}{|l|l|l|r|}
\hline
\rowcolor[HTML]{EFEFEF} 
\multicolumn{1}{|c|}{\cellcolor[HTML]{EFEFEF}\textbf{Software system coded construct}} & \multicolumn{1}{c|}{\cellcolor[HTML]{EFEFEF}\textbf{Description}} & \multicolumn{1}{c|}{\cellcolor[HTML]{EFEFEF}\textbf{Studies}} & \multicolumn{1}{c|}{\cellcolor[HTML]{EFEFEF}\textbf{Count}} \\ \hline
class assignments & \begin{tabular}[c]{@{}l@{}}The study used a software system for academic settings,\\ previously developed for some specific purpose (be \\ tested, refactored, ...) or developed during the condu-\\ ction of the study by students\end{tabular} & \begin{tabular}[c]{@{}l@{}}S92, S95, S96, S99, S102, S103, \\ S106, S113, S115, S117, S120, S123, \\ S125, S126, S133, S135, S136, S140, \\ S141, S144, S146, S147, S150, S151, \\ S158, S163, S169, S170, S171, S172, \\ S173, S174, S176, S177, S182, S193, \\ S196\end{tabular} & 37 \\ \hline
none & No software system was used in the study & \begin{tabular}[c]{@{}l@{}}S93, S104, S110, S112, S114, S118,\\ S119, S121, S122, S127, S129, S131, \\ S132, S137, S138, S139, S145, S153, \\ S154, S155, S159, S160, S161, S165, \\ S166, S178, S179, S181, S186, S187, \\ S189, S191, S195\end{tabular} & 33 \\ \hline
tool & \begin{tabular}[c]{@{}l@{}}The study used a machine learning/computational\\  intelligence, some kind of algorithm, gamecard, \\ or CASE tools\end{tabular} & \begin{tabular}[c]{@{}l@{}}S91, S94, S97, S98, S100, S101, S107,\\ S111, S124, S143, S148, S156, S175,\\ S185, S192\end{tabular} & 15 \\ \hline
software repository & \begin{tabular}[c]{@{}l@{}}The study used a software repository that stores so-\\ me kind of software artifacts, (e.g. code, documents, \\ or tasks)\end{tabular} & \begin{tabular}[c]{@{}l@{}}S116, S134, S149, S152, S164, S180,\\ S188, S194\end{tabular} & 8 \\ \hline
given software project & \begin{tabular}[c]{@{}l@{}}The study uses software (or requirements of it) that\\  has not yet been developed, but it was during the \\ conduction of the study\end{tabular} & \begin{tabular}[c]{@{}l@{}}S105, S108, S142, S157, S183, S184,\\ S190\end{tabular} & 7 \\ \hline
crowdsourcing projects & \begin{tabular}[c]{@{}l@{}}The study has crowdsourcing projects as a soft-\\ ware system\end{tabular} & S162, S167, S168 & 3 \\ \hline
\begin{tabular}[c]{@{}l@{}}implementation of a dataware\\ house system\end{tabular} & \begin{tabular}[c]{@{}l@{}}The study used data from implementation of a \\ dataware house system\end{tabular} & S109 & 1 \\ \hline
programming contest & \begin{tabular}[c]{@{}l@{}}The study used what is developed in a program-\\ ming contest\end{tabular} & S128 & 1 \\ \hline
industry projects & The study used data regarding industrial projects & S130 & 1 \\ \hline
\end{tabular}%
}
\end{table}

\newpage
   
 \bibliographystyle{elsarticle-harv} 
 \bibliography{zotero-refs}





\end{document}